\documentclass[showpacs,aps,graphicx,twocolumn]{revtex4}
\usepackage[tbtags]{amsmath}
\usepackage{mathrsfs}
\usepackage{amsfonts}
\usepackage{amssymb}
\usepackage{graphicx}
\usepackage{subfigure}
\usepackage{eufrak}
\usepackage{multirow}
\usepackage{float}
\usepackage[percent]{overpic}
\usepackage{bm}
\usepackage{xcolor}
\usepackage{blkarray}
\usepackage[colorlinks,linkcolor=blue,anchorcolor=blue,citecolor=blue,urlcolor=blue]{hyperref}
\usepackage{soul,color,xcolor}
\usepackage{amssymb}

\begin{document}

\title{Effective schemes for fusion of hyperentangled $W$ states}

\author{Wen-Xiu Zhang\textsuperscript{1}, Wen-Qiang Liu\textsuperscript{2}, and Hai-Rui Wei\textsuperscript{1,}}
\email[]{hrwei@ustb.edu.cn}

\address{\textsuperscript{\rm1} School of Mathematics and Physics, University of Science and Technology Beijing, Beijing 100083, China}

\address{\textsuperscript{\rm2} Department of Mathematics and Physics, Shijiazhuang Tiedao University, Shijiazhuang 050043, China}

\date{\today }

\begin{abstract}

Hyperentangled states are fascinating resources in quantum information processing as they can significantly increase the channel capacity and enhance noise resistance. 
We explore a hyperfusion mechanism to fuse one $n$-photon hyper-$W$ state and one $m$-photon hyper-$W$ state into a large-scale $(n+m-2)$-photon hyper-$W$ state. 
Another mechanism to fuse one $n$-photon hyper-$W$ state, one $m$-photon hyper-$W$ state, and one $t$-photon hyper-$W$ state into an $(n+m+t-3)$-photon hyper-$W$ state is also proposed. 
These two hyperfusion mechanisms are constructed employing only polarizing beam splitters, balanced beam splitters, half-wave plates, single-photon detectors, and cross-Kerr nonlinearities. 
Conditional quantum gates, path couplers, and ancillary photons are not required in our constructions. 
Moreover, our fused $W$ states are hyperentangled in the polarization and spatial degrees of freedom of single-photon systems. 
The presence of only one garbage output state demonstrates that high eﬃciency can be achieved in our schemes.

\end{abstract}


\maketitle

\section{Introduction}\label{sec1}

Quantum entanglement, which has nonclassical and nonlocal properties, is the foundation for many quantum computation and quantum communication tasks, such as one-way quantum computation \cite{one-way1,one-way2}, quantum key distribution \cite{QKD1,QKD2,QKD3},
quantum teleportation \cite{teleportation1,teleportation2,teleportation3,teleportation4},
quantum secure direct communication \cite{QSDC1,QSDC2,QSDC3,QSDC4},
entanglement purification \cite{purification1,purification2,purification3},
entanglement concentration \cite{concentration1,concentration2,concentration3},
quantum dense coding \cite{QDC1},
and quantum secret sharing \cite{QSS1,QSS2}.
Multiqubit entangled states are essential for measurement- or fusion-based quantum computing, distributed quantum computing, and quantum networks \cite{fusion-comput,Li1,Li2}.
There are different types of multiqubit entangled states, such as Greenberger-Horne-Zeilinger states, $W$ states, cluster states, and Dicke states; these are key resources for different quantum information processing (QIP) tasks.
Among these types of entangled state, $W$ states are intrinsically robust against qubit loss and can carry entanglement for longer times due to their web-like structures \cite{robust}. Hence, $W$ states are a fascinating resource in quantum information tasks, such as quantum key distribution \cite{W-QKD}, quantum teleportation \cite{teleportation3,W-teleportation1}, quantum metrology \cite{W-metrology}, quantum secure communication \cite{W-secure}, optimal universal quantum cloning machines \cite{W-clone},  quantum leaders \cite{W-leader}, anonymous quantum networks \cite{W-secure}, and quantum thermodynamics (e.g., thermalization machines) \cite{W-thermodynamics};
however, the preparation of large-scale $W$ states is challenging because the number of particles increases, the entanglement structure becomes more complex \cite{W-preparation}.
Therefore, it is desirable to find an effective way to generate large-scale $W$ state.

Expansion and fusion are two possible methods for preparing large-scale $W$ states.
In 2008, Tashima \emph{et al.} \cite{expand-2008} introduced an optical gate to expand an $m$-photon  polarization $W$ state to an $(m+2)$-photon $W$ state assisted by two 50:50 beam splitters (BSs), one phase shifter (PS), and one two-photon Fock state.
This scheme was experimentally demonstrated in 2010 \cite{expand-2010}.
Some improved $W$ state-expansion strategies have also been proposed.
In 2009, Tashima \emph{et al.} \cite{expand-2009} demonstrated the expansion of an $n$-photon $W$ state to an $(n+1)$-photon $W$ state using a polarization beam splitter (PBS) and one additional single photon with a defined polarization.
In 2011, Ikuta \emph{et al.} \cite{expand-2011} expanded an $n$-photon $W$ state to an $(n+m)$-photon $W$ state using  passive linear optics and one $m$-photon Fock state as ancillary resources.
Deterministic expansion approaches with the help of Toffoli gates or controlled-$H$ gates have also been proposed \cite{fusion-expansion-Toffoli,fusion-expansion-F}.
Compared to the expansion method, the fusion approach enables a substantially large increase in the size of a $W$ state.

The first fusion scheme was proposed by \"{O}zdemir \emph{et al.} \cite{fusion-NJP} to fuse an $n$-photon $W$ state and an $m$-photon $W$ state into an $(n+m-2)$-photon $W$ state using one PBS.
Many improved and interesting fusion protocols were subsequently proposed.
For example, to create large-scale $W$ states rapidly, strategies for fusing multiple $W$ states simultaneously have been proposed \cite{fusion-4W,fusion-Fredkin2,fusion-Zheng}.
Additionally, $W$ state-fusion strategies for photonic qubits \cite{fusion-NJP}, electronic qubits \cite{fusion-QD1,fusion-QD2}, atomic qubits \cite{fusion-atom1,fusion-atom2,fusion-atom3,fusion-atom4}, nitrogen-vacancy-center qubits \cite{fusion-NV}, and hybrid systems \cite{fusion-hybrid1,fusion-hybrid2} have been proposed to address different QIP tasks.
An interesting qubit-loss-free fusion mechanism was proposed to further increase fusion efficiency and decrease the number of fusion steps \cite{fusion-loss1,fusion-loss2,fusion-loss3,fusion-loss4}.
The success probability of a fused $W$ state has been enhanced by integrating conditional controlled gates (such as Fredkin gates, Toffoli gates, and CNOT gates) \cite{fusion-Fredkin1,fusion-Fredkin2} or enlarging Hilbert spaces \cite{fusion-space}.
The intrinsic probabilistic nature of a linear optical fused $W$ state has been overcome by introducing cross-Kerr nonlinearity \cite{fusion-Zheng};
however, the fusion of hyperentangled $W$ states remains largely unexplored, both theoretically and experimentally, despite the many remarkable advantages hyperentanglement offers in quantum communication and quantum computation.

Compared to normal entanglements in one degree of freedom (DOF) of quantum systems, hyperentanglements are entangled in multiple DOFs simultaneously.
Hyperentanglement offers access to several outstanding merits, such as boosting channel capacity, enhancing noise resistance, speeding up quantum computation, simplifying experimental setup, and reducing the use of quantum resources.
Moreover, hyperentanglement has strong potential for solving some sophisticated tasks that are intractable for single-DOF systems, such as complete Bell-state analysis with linear optics \cite{BSA1}.
Based on these unique features, hyperparallel quantum systems have led to a wide range of applications, such as
hyperswapping \cite{hyper-swapping1,hyper-swapping2},
hyper-Bell-states analysis \cite{hyper-BSA1,hyper-BSA2},
hyperentanglement purification \cite{hyper-purification1,hyper-purification2},
hyperparallel quantum computing \cite{hyper-computation1,hyper-computation2} and hyperentanglement concentration \cite{hyper-concentration1,hyper-concentration2}.
Photons are often regarded as natural carriers for hyperparallel operations, as hyperstates can be encoded in different qubit-like DOFs, such as polarization, path, frequency, orbital angular momentum, transverse momentum-position, multiple spatial modes, and time-bins \cite{Du1,Du2}.
Among  these DOFs, the spatial and polarization DOFs are most popular due to their relative ease of encoding and manipulation. Moreover, the spatial DOF is more robust.

In this paper, we present two practical hyper-$W$ state fusion schemes assisted by cross-Kerr mediums. Here, the hyperentanglement is encoded in the polarization and spatial DOFs of single-photon systems.
In the two-fusion scheme, an ($n+m-2$)-photon hyper-$W$ state can be obtained by fusing one $n$-photon hyper-$W$ state and with one $m$-photon hyper-$W$ state.
In the three-fusion scheme, an ($n+m+t-3$)-photon hyper-$W$ state can be obtained by fusing one $n$-photon hyper-$W$ state, one $m$-photon hyper-$W$ state, and one $t$-photon hyper-$W$ state.
Unlike standard constructions for single-DOF quantum systems, our schemes operate in a hyperparallel fashion.
Moreover, they are simple and compact because  path couplers, ancillary photons, and conditional controlled gates (i.e., CNOT, Toffoli, Fredkin, and partial-SWAP gates) are not required.
Most of the output states can be recycled for further fusion, enhancing resource utilization.

The remainder of this paper is organized as follows. 
In Sec. \ref{sec2}, we provide a scheme for fusing one $n$-photon hyper-$W$ state and one $m$-photon hyper-$W$ state into an $(n+m-2)$-photon hyper-$W$ state. 
In Sec. \ref{sec3}, we extend the scheme to fuse one $n$-photon hyper-$W$ state, one $m$-photon hyper-$W$ state, and one $t$-photon hyper-$W$ state into an $(n+m+t-2)$-photon hyper-$W$ state. 
In Sec. \ref{sec4}, we evaluate the performance of our schemes. 
Finally, in Sec. \ref{sec5}, we give a brief conclusion.

\section{two-fused hyperstates assisted by cross-Kerr nonlinearity } \label{sec2}

The construction we designed enables fusng one $n$-photon hyper-$W$ state and one $m$-photon hyper-$W$ state into an $(n+m-2)$-photon hyper-$W$ state.
Here, the hyperqubits are encoded in polarization and spatial states of single-photon systems, that is
$|H\rangle\equiv|\textbf{0}\rangle_P$,
$|V\rangle\equiv|\textbf{1}\rangle_P$,
$|a_0\rangle\equiv|\textbf{0}\rangle_S$,
$|a_1\rangle\equiv|\textbf{1}\rangle_S$,
$|b_0\rangle\equiv|\textbf{0}\rangle_S$, and
$|b_1\rangle\equiv|\textbf{1}\rangle_S$.
Here, $|H\rangle$ and $|V\rangle$ denote the horizontal and vertical polarization states, respectively.
Further, $|a_0\rangle$ and $|a_1\rangle$ ($|b_0\rangle$ and $|b_1\rangle$) denote the two spatial mode states, while
$|\textbf{0}\rangle$ and $|\textbf{1}\rangle$ are the logical qubit states.

\begin{figure*}[htpb]
\centering
\includegraphics[width=0.6\linewidth]{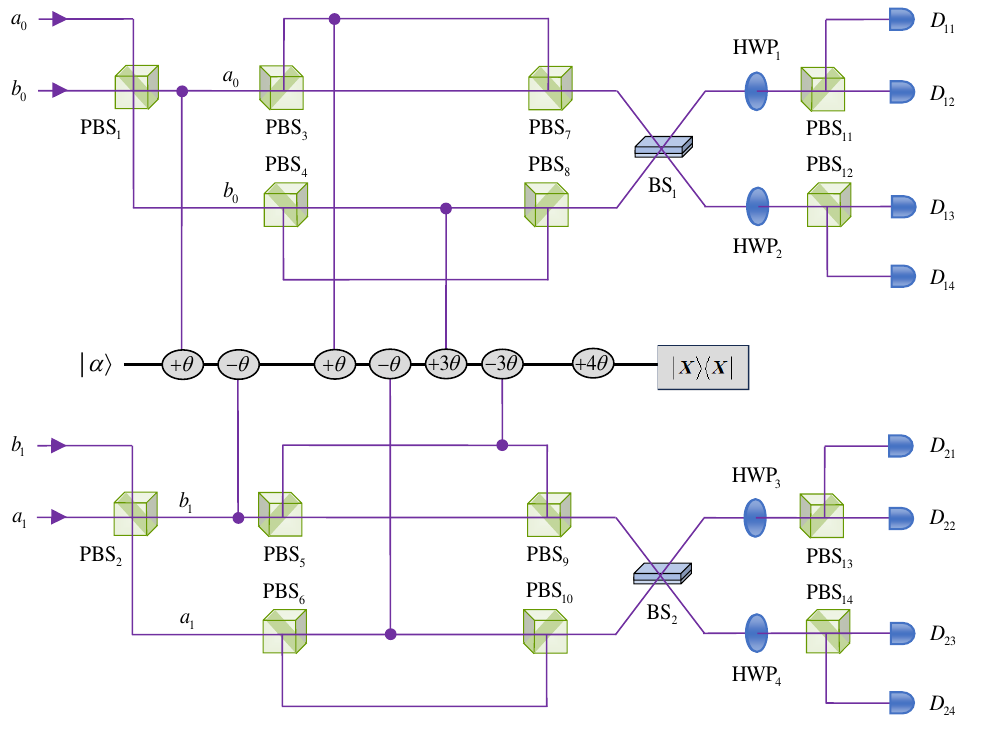}
\caption{Schematic diagram of the two-fusion protocol for fusing one $n$-photon hyper-$W$ and one $m$-photon hyper-$W$ state into an $(n+m-2)$-photon hyper-$W$ state.
A polarization beam splitter, $\textrm{PBS}_{i}$, transmits the horizontal polarization state and reflects the vertical polarization state, respectively.
A half wave plate, HWP$_j$, performs a polarization Hadamard operation on the incident photon.
A 50:50 beam splitter, BS$_k$, performs a spatial Hadamard operation on the incident photon.
Here, $D_{l}$ denotes the common single-photon detector and
$| X \rangle \langle X|$ represents an $X$ homodyne measurement.}
\label{figure1}
\end{figure*}

As shown in Fig. \ref{figure1}, two parties, Alice and Bob, possess an $n$-photon polarization-spatial hyper-W state $|W_n\rangle$ and an $m$-photon polarization-spatial hyper-W state $|W_m\rangle$ ($n, m \ge 2$), respectively.
Here,
\begin{eqnarray}\label{eq1}
\begin{split}
|W_n\rangle=\; & \frac{1}{ n } \big[|(n-1)_H\rangle_P     |1_V\rangle_P     \\& + \sqrt {n-1} |W_{n-1} \rangle_P |1_H\rangle_P \big]\\&
                       \otimes \big[|(n-1)_{a_0}\rangle_S |1_{a_1}\rangle_S \\& + \sqrt{n-1}  |W_{n-1}\rangle_S  |1_{a_0}\rangle_S \big],\\
|W_m\rangle=\; & \frac{1}{ m } \big[|(m-1)_H\rangle_P     |1_V\rangle_P     \\& + \sqrt{m-1}  |W_{m-1} \rangle_P |1_H\rangle_P \big]\\&
                       \otimes \big[|(m-1)_{b_0}\rangle_S |1_{b_1}\rangle_S \\& + \sqrt{m-1}  |W_{m-1}\rangle_S  |1_{b_0}\rangle_S \big],
\end{split}	
\end{eqnarray}
where $|(n-1)_H\rangle_P |1_V\rangle_P$  indicates that $(n-1)$ qubits are in the $H$-polarization state and one qubit is in the $V$-polarization state;
      $|W_{n-1}\rangle_P|1_H\rangle_P$  indicates that $(n-1)$ qubits are in the polarization $W$ state and  one qubit is in the $H$-polarization state;
      $|(n-1)_{a_0}\rangle_S |1_{a_1}\rangle_S$  indicates that $(n-1)$ qubits are in the spatial state $|\textbf{0}\rangle_S$ and one qubit is in the spatial state $|\textbf{1}\rangle_S$;
      $|W_{n-1}\rangle_S  |1_{a_0}\rangle_S$   indicates that $(n-1)$ qubits are in the spatial $W$ state and one qubit is in the spatial state $|\textbf{0}\rangle_S$.

Note that in the construction process, only the $n$th photon on Alice's side and the $m$th photon on Bob's side are sent to the fusion mechanism.
Now, let us introduce the procedure of our hyperfusion mechanism assisted by a cross-Kerr medium, step by step.

In the first step, the photon emitted from spatial $a_0$ ($a_1$) mixes with the photon emitted from spatial $b_0$ ($b_1$) at $\rm{PBS}_{1}$ ($\rm{PBS}_{2}$). Here, the PBS transmits the $H$-polarized component and reflects the $V$-polarized component. Subsequently, cross-Kerr medium introduces phase shifts $e^{\text{i}\theta}$ and $e^{-\text{i}\theta}$ to the components emitted from spatial $a_0$  and $a_1$, respectively.

In the second step, the cross-Kerr medium
introduces a phase shift $e^{\text{i}\theta}$   to the $V$ components emitted from spatial $a_0$ by the block composed of PBS$_3$ and PBS$_7$;
              introduces $e^{-\text{i}\theta}$  to the $H$ components emitted from spatial $a_1$ by the block composed of PBS$_6$ and PBS$_{10}$;
              introduces $e^{\text{i}3\theta}$  to the $H$ components emitted from spatial $b_0$ by the block composed of PBS$_4$ and PBS$_8$; and
             introduces $e^{-\text{i}3\theta}$ to the $V$ components emitted from spatial $b_1$ by the block composed of PBS$_5$ and PBS$_9$.
Then, an $X$-quadrature homodyne measurement is performed on the coherent state.
The detailed detection results and the corresponding quantum states of the system are given in the Appendix.

In the third step, two spatial Hadamard operations are performed on the incident photons using the 50:50 BS$_1$ and the 50:50 BS$_2$, and then four
polarization Hadamard operations are performed on the incident photons using half-wave plates $\text{HWP}_1$, $\text{HWP}_2$, $\text{HWP}_3$, and $\text{HWP}_4$.
Here, the 50:50 BS completes the transformations
\begin{eqnarray}\label{eq12}
\begin{split}
&a_{\text{in}}^\dag \xrightarrow{\text{BS}} \frac{1}{\sqrt 2}(a_{\text{out}}^\dag + b_{\text{out}}^\dag),\\
&b_{\text{in}}^\dag \xrightarrow{\text{BS}} \frac{1}{\sqrt 2}(a_{\text{out}}^\dag - b_{\text{out}}^\dag).
\end{split}
\end{eqnarray}
and the HWP completes the transformations
\begin{eqnarray}\label{eq13}
\begin{split}
&|H\rangle \xrightarrow{\text{HWP}} \frac{1}{\sqrt 2}(|H\rangle + |V\rangle),\\
&|V\rangle \xrightarrow{\text{HWP}} \frac{1}{\sqrt 2}(|H\rangle - |V\rangle).
\end{split}
\end{eqnarray}

After the click of the single-photon detectors, the corresponding feed-forward operations (see Tab. \ref{table1}) applied on the photons held by Alice or Bob.
This allows us to obtain various states.
Specifically, we may obtain the garbage state
\begin{eqnarray}\label{eq14}
\begin{split}
|\Phi_{e^{\text{i}0\theta}}\rangle \rightarrow |\Phi_{\text{F}}\rangle =\;& |(n-1)_H\rangle_P |(m-1)_H\rangle_P \\& \otimes |(n-1)_{a_0}\rangle_S |(m-1)_{b_0}\rangle_S,
\end{split}
\end{eqnarray}
with a probability of $\frac{1}{n^2 m^2}$. In this case, the scheme ends in a failure, producing a garbage state. Such event is called failure.

We may obtain the large-scale polarization state
\begin{eqnarray}\label{eq15}
\begin{split}
|\Phi_{e^{\text{i}\theta}}\rangle \rightarrow |\Phi_{\text{PS}^\text{p}}\rangle =\;& |W_{n+m-2}\rangle_P |(n-1)_{a_0}\rangle_S \\& \otimes |(m-1)_{b_0}\rangle_S,
\end{split}
\end{eqnarray}
with a probability of $\frac{n+m-2}{n^2 m^2}$. Such an event is called partially successful.

We may obtain the partially recyclable polarization state
\begin{eqnarray}\label{eq16}
\begin{split}
|\Phi_{e^{\text{i}2\theta}}\rangle \rightarrow |\Phi_{\text{PR}^\text{pp}}\rangle =\;& |W_{n-1}\rangle_P |W_{m-1}\rangle_P \\& \otimes |(n-1)_{a_0}\rangle_S |(m-1)_{b_0}\rangle_S,
\end{split}
\end{eqnarray}
with a probability of $\frac{(n-1)(m-1)}{n^2 m^2}$. Here, the short $W $state $|W_{n-1}\rangle_P \otimes |W_{m-1}\rangle_P$ can be recycled. Thus, such an event is called partially recyclable.

We may obtain the large-scale spatial state
\begin{eqnarray}\label{eq17}
\begin{split}
|\Phi_{e^{\text{i}3\theta}}\rangle \rightarrow |\Phi_{\text{PS}^\text{s}}\rangle =\;& |(n-1)_H\rangle_P |(m-1)_H\rangle_P \\& \otimes |W_{n+m-2}\rangle_S,
\end{split}
\end{eqnarray}
with a probability of $\frac{n+m-2}{n^2 m^2}$.

We may obtain the desired large-scale polarization-spatial hyperstate
\begin{eqnarray}\label{eq18}
\begin{split}
|\Phi_{e^{\text{i}4\theta}}\rangle \rightarrow |\Phi_{\text{S}}\rangle = |W_{n+m-2}\rangle_P |W_{n+m-2}\rangle_S,
\end{split}
\end{eqnarray}
with a probability of $\frac{(n+m-2)^2}{n^2 m^2}$. Such an event is called successful.

We may obtain the partially recyclable polarization state and the large-scale spatial state
\begin{eqnarray}\label{eq19}
\begin{split}
|\Phi_{e^{\text{i}5\theta}}\rangle \rightarrow |\Phi_{\text{PR}^{\text{pp}},\text{PS}^{\text{s}}}\rangle =\;& |W_{n-1}\rangle_P |W_{m-1}\rangle_P \\& \otimes |W_{n+m-2}\rangle_S,
\end{split}
\end{eqnarray}
with a probability of $\frac{(n-1)(m-1)(n+m-2)}{n^2 m^2}$.

We may obtain the partially recyclable spatial state
\begin{eqnarray}\label{eq20}
\begin{split}
|\Phi_{e^{\text{i}6\theta}}\rangle \rightarrow |\Phi_{\text{PR}^{\text{ss}}}\rangle =\;& |(n-1)_H\rangle_P |(m-1)_H\rangle_P \\& \otimes |W_{n-1}\rangle_S |W_{m-1}\rangle_S,
\end{split}
\end{eqnarray}
with a probability of $\frac{(n-1)(m-1)}{n^2 m^2}$.

We may obtain the large-scale polarization state and the partially recyclable spatial state
\begin{eqnarray}\label{eq21}
\begin{split}
|\Phi_{e^{\text{i}7\theta}}\rangle \rightarrow |\Phi_{\text{PS}^{\text{p}},\text{PR}^{\text{ss}}}\rangle =\;& |W_{n+m-2}\rangle_P |W_{n-1}\rangle_S \\& \otimes |W_{m-1}\rangle_S,
\end{split}
\end{eqnarray}
with a probability of $\frac{(n+m-2)(n-1)(m-1)}{n^2 m^2}$.

We may obtain the partially recyclable polarization state and the partially recyclable spatial state
\begin{eqnarray}\label{eq22}
\begin{split}
|\Phi_{e^{\text{i}8\theta}}\rangle \rightarrow |\Phi_{\text{PR}^{\text{ppss}}}\rangle =\;& |W_{n-1}\rangle_P |W_{m-1}\rangle_P |W_{n-1}\rangle_S \\& \otimes |W_{m-1}\rangle_S,
\end{split}
\end{eqnarray}
with a probability of $\frac{(n-1)^2(m-1)^2}{n^2 m^2}$.

\begin{table*}[htbp]
\centering
\caption{Relations between the quantum nondemolotion detection (QND), the triggers of the single-photon detectors, and the classical feed-forward operations used to complete the two-fusion scheme.
Here, $Z_{\text{Alice/Bob}}^\text{S}$ denotes the spatial feed-forward operation, $|a_0\rangle_S \langle a_0|- |a_1 \rangle_S\langle a_1|$ applied to  the photons held by Alice and Bob, and 
$Z^\text{P}_{\text{Alice/Bob}}$ denotes the polarization feed-forward operation, $|H\rangle_P\langle H|-|V\rangle_P\langle V|$ applied to the photons held by Alice and Bob.}	\label{table1}
\begin{tabular}{ccccc}	

\hline 
\hline

$|X\rangle\langle X|$ & Single-photon detector pairs & Feed-forward operations  & Result   & Success probability \\	

\hline
	
\multirow{3}{*}{$|\alpha \rangle$}  & $({D_{23}},{D_{23}}), ({D_{23}},{D_{24}}), ({D_{24}},{D_{24}})$   & \multirow{3}{*}{None} & \multirow{3}{*}{$|\Phi_{\rm{F}}\rangle $}
                                    & \multirow{3}{*}{$\dfrac{1}{{n^2}{m^2}}$} \\
                                    & $({D_{23}},{D_{22}}), ({D_{23}},{D_{21}}), ({D_{21}},{D_{23}})$ & & & \\
                                    & $({D_{22}},{D_{22}}), ({D_{22}},{D_{21}}), ({D_{21}},{D_{21}})$ & & & \\

     & & & &\\

\multirow{2}{*}{$|e^{\textrm{i}\theta}\alpha\rangle$}  & $(D_{23},D_{23}), (D_{24},D_{24}), (D_{22},D_{22})$   &     None  &  \multirow{2}{*}{$|\Phi_{\text{PS}^\text{p}}\rangle$}
                                                       & \multirow{2}{*}{$\dfrac{n+m-2}{n^2 m^2}$} \\                                                          
                                                       & $(D_{21},D_{21}),(D_{22},D_{23}),(D_{21},D_{24})$   & $Z_\text{Alice}^{\text{P}}$ & & \\

     & & & &\\

\multirow{2}{*}{$|e^{\textrm{i}2\theta}\alpha\rangle$}  & $({D_{23}},{D_{23}}), ({D_{23}},{D_{24}}), ({D_{24}},{D_{24}})$  & \multirow{2}{*}{None}  
                                                        & \multirow{2}{*}{$|\Phi_{\text{PR}^\text{PP}}\rangle$}          & \multirow{2}{*}{$\dfrac{(n-1)(m-1)}{n^2 m^2}$ } \\
                                                        & $({D_{22}},{D_{22}}), ({D_{21}},{D_{22}}), ({D_{21}},{D_{21}})$  &   &  & \\	

     & & & &\\

\multirow{5}{*}{$|e^{\textrm{i}3\theta}\alpha\rangle$}  & $({D_{12}},{D_{23}}), ({D_{11}},{D_{23}}), ({D_{12}},{D_{24}})$  & \multirow{2}{*}{None}  
                                                        & \multirow{5}{*}{$|\Phi_{\text{PS}^\text{s}}\rangle$}           & \multirow{5}{*}{$\dfrac{n+m-2}{n^2 m^2}$ } \\
                                                        & $({D_{11}},{D_{24}}), ({D_{14}},{D_{22}}), ({D_{13}},{D_{21}}),({D_{14}},{D_{21}})$  &   &  & \\
                                                        & $({D_{13}},{D_{23}}), ({D_{14}},{D_{23}}), ({D_{13}},{D_{24}})$  & \multirow{3}{*}{$Z_\text{Alice}^{\text{S}}$}  &  & \\	
                                                        & $({D_{14}},{D_{24}}), ({D_{12}},{D_{22}}), ({D_{11}},{D_{22}})$  &   &  & \\
                                                        & $({D_{12}},{D_{21}}), ({D_{11}},{D_{21}}), ({D_{13}},{D_{22}})$  &   &  &\\		
                                                            
     & & & &\\

\multirow{4}{*}{$|e^{\textrm{i}4\theta}\alpha\rangle$}  & $({D_{12}},{D_{23}}),({D_{11}},{D_{24}}),({D_{13}},{D_{22}}),({D_{14}},{D_{21}})$  & None  
                                                        & \multirow{4}{*}{$|\Phi_{\text{S}}\rangle$}           & \multirow{4}{*}{$\dfrac{(n+m-2)^2}{n^2 m^2}$} \\
                                                        & $({D_{11}},{D_{23}}), ({D_{12}},{D_{24}}), ({D_{14}},{D_{22}}), ({D_{13}},{D_{21}})$  & $Z_\text{Alice}^\text{P},Z_\text{Alice}^\text{S}$ 
                                                        &  & \\
                                                        & $({D_{13}},{D_{23}}), ({D_{14}},{D_{24}}), ({D_{12}},{D_{22}}),({D_{11}}, {D_{21}})$  & $Z_\text{Alice}^\text{P}$  &  & \\	
                                                        & $({D_{14}},{D_{23}}), ({D_{13}},{D_{24}}), ({D_{11}},{D_{22}}),({D_{12}}, {D_{21}})$  & $Z_\text{Alice}^\text{S}$  &  & \\

     & & & &\\

\multirow{4}{*}{$|e^{\textrm{i}5\theta}\alpha\rangle$}  & $({D_{12}},{D_{23}}),({D_{11}}, {D_{23}}),({D_{12}}, {D_{24}}),({D_{11}},{D_{24}})$  & \multirow{2}{*}{None}  
                                                        & \multirow{4}{*}{$|\Phi_{\text{PR}^{\text{pp}},\text{PS}^{\text{s}}}\rangle$}       &  \multirow{4}{*}{$\dfrac{(n-1)(m-1)(n+m-2)}{n^2 m^2}$}\\
                                                        
                                                        & $({D_{13}},{D_{22}}), ({D_{14}},{D_{22}}),({D_{13}}, {D_{21}}),({D_{14}},{D_{21}})$  &   &  & \\
                                                        
                                                        & $({D_{13}},{D_{23}}), ({D_{14}},{D_{23}}),({D_{13}}, {D_{24}}),({D_{14}},{D_{24}})$  & \multirow{2}{*}{$Z_\text{Bob}^\text{P}$}  &  & \\	
                                                        
                                                        & $({D_{12}},{D_{22}}), ({D_{11}},{D_{22}}),({D_{12}}, {D_{21}}),({D_{11}},{D_{21}})$  &  &  & \\

     & & & &\\

\multirow{2}{*}{$|e^{\textrm{i}6\theta}\alpha\rangle$}  & $({D_{12}},{D_{12}}), ({D_{11}},{D_{12}}), ({D_{11}},{D_{11}})$  & \multirow{2}{*}{None}  
                                                        & \multirow{2}{*}{$|\Phi_{\text{PR}^{\text{ss}}}\rangle$}        & \multirow{2}{*}{$\dfrac{(n-1)(m-1)}{n^2 m^2}$ } \\
                                                        & $({D_{13}},{D_{13}}),({D_{13}},{D_{14}}),({D_{14}},{D_{14}})$  &   &  & \\

     & & & &\\	
     
\multirow{2}{*}{$|e^{\textrm{i}7\theta}\alpha\rangle$}  &$({D_{11}},{D_{11}}),({D_{12}},{D_{12}}), ({D_{13}},{D_{13}}), ({D_{14}},{D_{14}})$ & None  
                                                        & \multirow{2}{*}{$|\Phi_{\text{PS}^{\text{p}},\text{PR}^{\text{ss}}}\rangle$}    & \multirow{2}{*}{$\dfrac{(n-1)(m-1)(n+m-2)}{n^2 m^2}$} \\
                                                        & $({D_{12}},{D_{13}}),({D_{11}},{D_{14}}))$                                      & $Z_\text{Alice}^\text{P}$  &  & \\

     & & & &\\	
     
\multirow{2}{*}{$|{{e^{i8\theta}}\alpha}\rangle$}  & $({D_{12}},{D_{12}}), ({D_{11}},{D_{12}}), ({D_{11}},{D_{11}})$  & \multirow{2}{*}{None}  
                                                   & \multirow{2}{*}{$|\Phi_{\text{PR}^{\text{ppss}}}\rangle$}      & \multirow{2}{*}{$\dfrac{(n-1)^2 (m-1)^2}{n^2 m^2}$} \\
                                                   & $({D_{13}},{D_{13}}), ({D_{13}},{D_{14}}), ({D_{14}},{D_{14}})$  &   &  & \\	

\hline
\hline
	\end{tabular}
\end{table*}

\section{Three-fused hyper-W states assisted by cross-Kerr nonlinearities }\label{sec3}

In this section, we generalize the two-fusion scheme described in Sec. \ref{sec2} to the three-fusion case. That is, the scheme can fuse one $n$-photon hyper-$W$ state, one $m$-photon hyper-$W$ state, and one $t$-photon hyper-$W$ state into an $(n+m+t-3)$-photon hyper-$W$ state. The initial states possessed by Alice, Bob, and Charlie are given by
\begin{eqnarray}\label{eq23}
\begin{split}
  |W_n\rangle =\;& \frac{1}{ n } \big[|(n-1)_H\rangle_P     |1_V\rangle_P     \\& + \sqrt {n-1}|W_{n-1} \rangle_P |1_H\rangle_P \big]\\&
                         \otimes \big[|(n-1)_{a_0}\rangle_S |1_{a_1}\rangle_S \\& + \sqrt{n-1} |W_{n-1}\rangle_S  |1_{a_0}\rangle_S \big],\\
  |W_m\rangle =\;& \frac{1}{ m } \big[|(m-1)_H\rangle_P     |1_V\rangle_P     \\& + \sqrt{m-1} |W_{m-1} \rangle_P |1_H\rangle_P ]\\&
                         \otimes \big[|(m-1)_{b_0}\rangle_S |1_{b_1}\rangle_S \\& + \sqrt{m-1} |W_{m-1}\rangle_S  |1_{b_0}\rangle_S \big],\\
  |W_t\rangle =\;& \frac{1}{ t } \big[|(t-1)_H\rangle_P     |1_V\rangle_P     \\& + \sqrt{t-1} |W_{t-1} \rangle_P |1_H\rangle_P \big]\\&
                         \otimes \big[|(t-1)_{c_0}\rangle_S |1_{c_1}\rangle_S \\& + \sqrt{t-1} |W_{t-1}\rangle_S  |1_{c_0}\rangle_S \big].
\end{split}	
\end{eqnarray}
Here,
$|H\rangle  \equiv|\textbf{0}\rangle_P$,       $|V\rangle\equiv|\textbf{1}\rangle_P$,
$|a_0\rangle\equiv|\textbf{0}\rangle_S$,     $|a_1\rangle\equiv|\textbf{1}\rangle_S$,
$|b_0\rangle\equiv|\textbf{0}\rangle_S$,     $|b_1\rangle\equiv|\textbf{1}\rangle_S$,
$|c_0\rangle\equiv|\textbf{0}\rangle_S$, and $|c_1\rangle\equiv|\textbf{1}\rangle_S$.

\begin{figure*}
	\centering
	\includegraphics[width=1\linewidth]{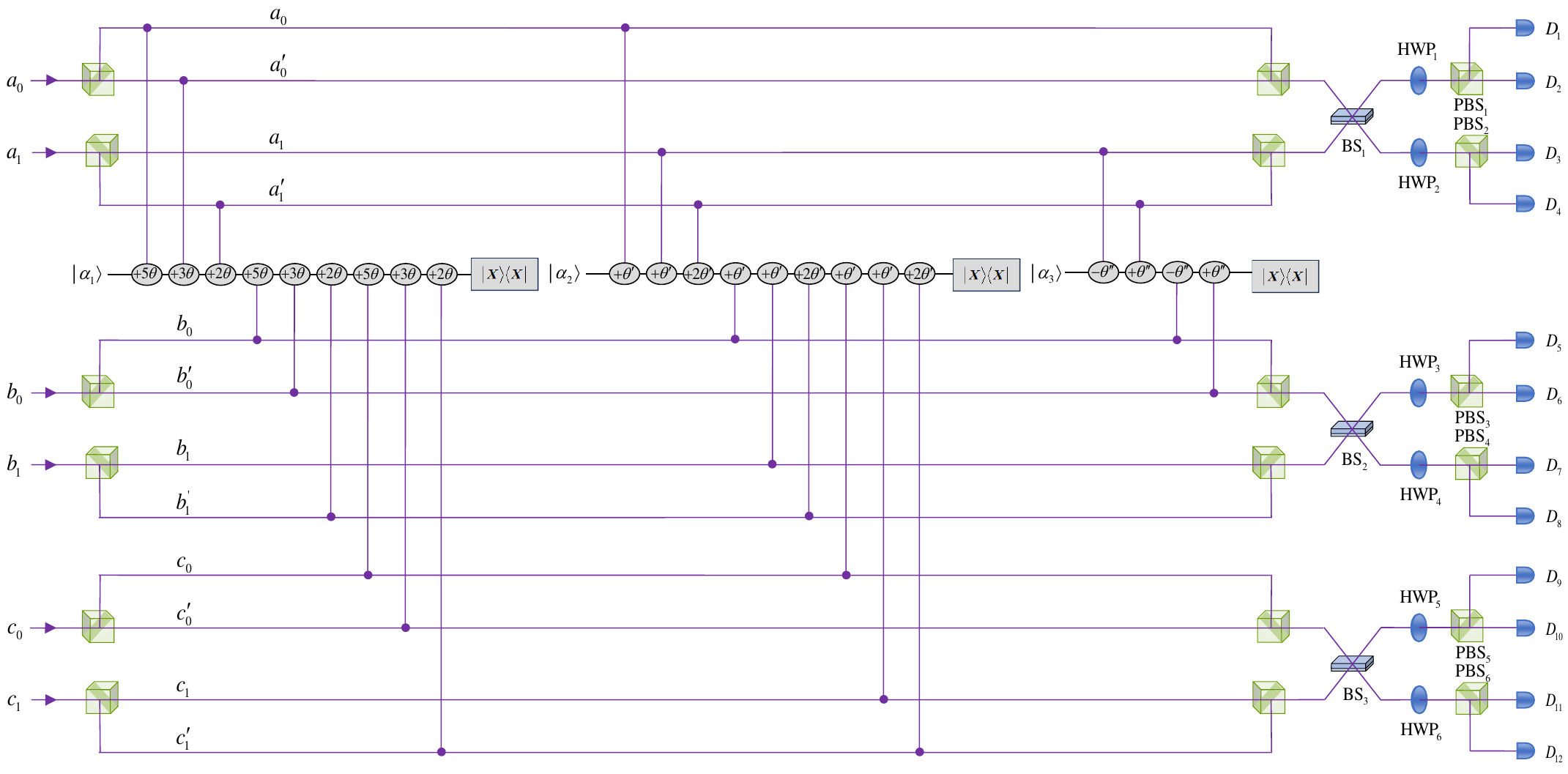}
	\caption{Schematic diagram of the three-fusion protocol for fusing one $n$-photon hyper-$W$ state, one $m$-photon hyper-$W$ state, and one $t$-photon hyper-$W$ state into an $(n+m+t-3)$-photon hyper-$W$ state.}
	\label{figure2}
\end{figure*}

As shown in Fig. \ref{figure2}, the three incident photons interact with the coherent states $|\alpha _1\rangle$, $|\alpha _2\rangle$, and $|\alpha _3\rangle$ in succession. They then undergo from polarization Hadamard operations, spatial Hadamard operations, and single-photon detections in succession.
As shown in Tabs. \ref{table2}-\ref{table3}, based on the measurement results of the three coherent states and the triggers of the single-photon detectors, we can obtain different types of output state.

In detail, if the probe mode receives phase shifts $(6\theta, 6\theta^{\prime}, \theta^{\prime\prime})$, we obtain the garbage state
\begin{eqnarray}\label{eq24}
\begin{split}
|\Psi_\text{F}\rangle =\;& |(n-1)_H\rangle_P |(m-1)_H\rangle_P |(t-1)_H\rangle_P  \\& \otimes |(n-1)_{a_0}\rangle_S |(m-1)_{b_0}\rangle_S |(t-1)_{c_0}\rangle_S,
\end{split}
\end{eqnarray}
with a probability of $\frac{1}{n^2 m^2 t^2}$.

If the probe mode receives phase shifts $(7\theta, 4\theta^{\prime}, \pm2\theta^{\prime\prime})$, $(7\theta, 4\theta^{\prime}, 0\theta^{\prime\prime})$,
                                       or  $(7\theta, 4\theta^{\prime}, \pm\theta ^{\prime\prime})$,
we obtain the desired large-scale hyper-$W$ state
\begin{eqnarray}\label{eq25}
\begin{split}
|\Psi_{\text{S}}\rangle = |W_{n+m+t-3}\rangle_P |W_{n+m+t-3}\rangle_S,
\end{split}
\end{eqnarray}
with a success probability of $\frac{(n+m+t-3)^2}{n^2 m^2 t^2}$.

If the probe mode receives phase shifts $(9\theta, 5\theta^{\prime}, 0\theta^{\prime\prime})$, we  obtain the large-scale spatial state
\begin{eqnarray}\label{eq26}
\begin{split}
|\Psi_{\text{PS}^\text{s}}\rangle =\; & |(n-1)_H\rangle_P |(m-1)_H\rangle_P |(t-1)_H\rangle_P  \\& \otimes |W_{n+m-2}\rangle_S |(t-1)_{c_0}\rangle_S,
\end{split}
\end{eqnarray}
with a probability of $\frac{n+m-2}{n^2 m^2 t^2}$.

If the probe mode receives phase shifts $(12\theta, 4\theta^{\prime}, 0\theta^{\prime\prime})$,  we obtain the large-scale spatial state and the partially recyclable spatial state
\begin{eqnarray}\label{eq27}
\begin{split}
|\Psi_{\text{PS}^\text{s},\text{PR}^\text{s}}\rangle =\;& |(n-1)_H\rangle_P |(m-1)_H\rangle_P |(t-1)_H\rangle_P  \\& \otimes |W_{n+m-2}\rangle_S |W_{t-1}\rangle_S,
\end{split}
\end{eqnarray}
with a probability of $\frac{(n+m-2)(t-1)}{n^2 m^2 t^2}$.

If the probe mode receives phase shifts $(3\theta, 2\theta^{\prime}, 0\theta^{\prime\prime})$,  we obtain the partially recyclable polarization state and the large-scale spatial state
\begin{eqnarray}\label{eq28}
\begin{split}
|\Psi_{\text{PR}^\text{ppp},\text{PS}^\text{s}}\rangle =\;& |W_{n-1}\rangle_P |W_{m-1}\rangle_P |W_{t-1}\rangle_P  \\& \otimes |(t-1)_{c_0}\rangle_S |W_{n+m-2}\rangle_S,
\end{split}
\end{eqnarray}
with a probability of $\frac{(n-1)(m-1)(t-1)(n+m-2)}{n^2 m^2 t^2}$.

If the probe mode receives phase shifts $(6\theta, \theta^{\prime}, 0\theta^{\prime\prime})$, we obtain the partially recyclable polarization state, the large-scale spatial state, and the partially recyclable spatial state
\begin{eqnarray}\label{eq29}
\begin{split}
|\Psi_{\text{PR}^\text{ppp},\text{PS}^\text{s},\text{PR}^\text{s}}\rangle =\;& |W_{n-1}\rangle_P |W_{m-1}\rangle_P |W_{t-1}\rangle_P  \\& \otimes |W_{n+m-2}\rangle_S |W_{t-1}\rangle_S,
\end{split}
\end{eqnarray}
with a probability of $\frac{(n-1)(m-1)(t-1)^2(n+m-2)}{n^2 m^2 t^2}$.

If the probe mode receives phase shifts $(10\theta, 3\theta^{\prime}, \pm2\theta^{\prime\prime})$, we obtain the large-scale polarization state and the partially recyclable spatial state
\begin{eqnarray}\label{eq30}
\begin{split}
|\Psi_{\text{PS}^\text{p},\text{PR}^\text{ss}}\rangle =\;& |W_{n+m-2}\rangle_P |(t-1)_H\rangle_P  \\& \otimes |(n-1)_{a_0}\rangle_S |W_{m-1}\rangle_S  |W_{t-1}\rangle_S,
\end{split}
\end{eqnarray}
with a probability of $\frac{(n+m-2)(m-1)(t-1)}{n^2 m^2 t^2}$.

If the probe mode receives phase shifts $(12\theta, 2\theta^{\prime}, \pm\theta^{\prime\prime})$, we  obtain the large-scale polarization state and the partially recyclable spatial state
\begin{eqnarray}\label{eq31}
\begin{split}
|\Psi_{\text{PS}^\text{p},\text{PR}^\text{sss}}\rangle =\;& |W_{n+m-2}\rangle_P |(t-1)_H\rangle_P  \\& \otimes |W_{n-1}\rangle_S |W_{m-1}\rangle_S |W_{t-1}\rangle_S,
\end{split}
\end{eqnarray}
with a probability of $\frac{(n+m-2)(n-1)(m-1)(t-1)}{n^2 m^2 t^2}$.

If the probe mode receives phase shifts $(5\theta, 3\theta^{\prime}, \pm 2\theta^{\prime\prime})$, we  obtain the large-scale polarization state, the partially recyclable polarization spatial state, and partially recyclable spatial state
\begin{eqnarray}\label{eq32}
\begin{split}
|\Psi_{\text{PS}^\text{p},\text{PR}^\text{ps}}\rangle =\;& |W_{n+m-2}\rangle_P |W_{t-1}\rangle_P  \\& \otimes |(n-1)_{a_0}\rangle_S |W_{m-1}\rangle_S |(t-1)_{c_0}\rangle_S,
\end{split}
\end{eqnarray}
with a probability of $\frac{(n+m-2)(t-1)(m-1)}{n^2 m^2 t^2}$.

If the probe mode receives phase shifts $(8\theta, 2\theta^{\prime}, \pm\theta^{\prime\prime})$, we obtain the large-scale polarization state, the partially recyclable polarization state, and the partially recyclable spatial state
\begin{eqnarray}\label{eq33}
\begin{split}
|\Psi_{\text{PS}^\text{p},\text{PR}^\text{pss}}\rangle =\;& |W_{n+m-2}\rangle_P |W_{t-1}\rangle_P  \\& \otimes |W_{n-1}\rangle_S |W_{m-1}\rangle_S |(t-1)_{c_0}\rangle_S,
\end{split}
\end{eqnarray}
with a probability of $\frac{(n+m-2)(t-1)(n-1)(m-1)}{n^2 m^2 t^2}$.

If the probe mode receives phase shifts $(8\theta, 2\theta^{\prime}, \pm2\theta^{\prime\prime})$, we obtain the large-scale polarization state, the partially recyclable polarization state, and the partially recyclable spatial state
\begin{eqnarray}\label{eq34}
\begin{split}
|\Psi_{\text{PS}^\text{p},\text{PR}^\text{pss}}\rangle =\;& |W_{n+m-2}\rangle_P |W_{t-1}\rangle_P  \\& \otimes |(n-1)_{a_0}\rangle_S |W_{m-1}\rangle_S |W_{t-1}\rangle_S,
\end{split}
\end{eqnarray}
with a probability of $\frac{(n+m-2)(t-1)^2(m-1)}{n^2 m^2 t^2}$.

\begin{table*}[htpb]
	\centering
	\caption{Relations between the QND, the triggers of the single-photon detectors, and the classical feed-forward
operations used to complete the three-fusion scheme., including both the garbage outcomes and the desired large-scale hyper-$W$ state cases.}	\label{table2}
	\begin{tabular}{cccc}	
		\hline 	
		\hline

$|X\rangle\langle X|$  &  Single-photon detector pairs & Feed-forward operations & Result  \\
		
		\hline
	
\multirow{8}{*}{$|e^{6\textrm{i}\theta}\alpha _1\rangle |e^{6\textrm{i}\theta^{\prime}}\alpha _2\rangle |e^{\textrm{i}\theta^{\prime\prime}}\alpha _3\rangle$}
        &  $({D_1},{D_6},{D_{10}}), ({D_1},{D_6},{D_9}), ({D_1},{D_5},{D_{10}}),({D_1},{D_5},{D_9})$     
        & \multirow{8}{*}{\rm{None}} 
        & \multirow{8}{*}{$|\Psi_{\rm{F}}\rangle $} \\
        
        & $({D_1},{D_6},{D_{11}}), ({D_1},{D_6},{D_{12}}), ({D_1}, {D_5}, {D_{11}}), ({D_1},{D_5},{D_{12}})$ & & \\
        & $({D_1},{D_7},{D_{10}}), ({D_1},{D_7},{D_9}),    ({D_1}, {D_8}, {D_{10}}), ({D_1},{D_8},{D_9})$    & & \\
        & $({D_1},{D_7},{D_{11}}), ({D_1},{D_7},{D_{12}}), ({D_1}, {D_8}, {D_{11}}), ({D_1},{D_8},{D_{12}})$ & & \\
        & $({D_4},{D_6},{D_{10}}), ({D_4},{D_6},{D_9}),    ({D_4}, {D_5}, {D_{10}}), ({D_4},{D_5},{D_9})$    & & \\
        & $({D_4},{D_6},{D_{11}}), ({D_4},{D_6},{D_{12}}), ({D_4}, {D_5}, {D_{11}}), ({D_4},{D_5},{D_{12}})$ & & \\
        & $({D_4},{D_7},{D_{10}}), ({D_4},{D_7},{D_9}),    ({D_4}, {D_8}, {D_{10}}), ({D_4},{D_8},{D_9})$    & & \\
        &$({D_4},{D_7},{D_{11}}),  ({D_4},{D_7},{D_{12}}), ({D_4}, {D_8}, {D_{11}}), ({D_4},{D_8},{D_{12}})$ & &\\
       
        & & &\\

\multirow{16}{*}{$|e^{7\textrm{i}\theta}\alpha _1\rangle |e^{4\textrm{i}\theta ^{\prime}}\alpha_2\rangle |\alpha_3\rangle$}	                                     
        & $({D_2},{D_6},{D_{10}}), ({D_1},{D_5},{D_9}), ({D_3},{D_7},{D_{11}}), ({D_4},{D_8},{D_{12}})$ 
        & \rm{None} 
        & \multirow{16}{*}{$| \Psi_{\rm{S}} \rangle $} \\        
      
\multirow{16}{*}{$|e^{7\textrm{i}\theta}\alpha _1\rangle |e^{4\textrm{i}\theta ^{\prime}}\alpha_2\rangle |e^{\pm\textrm{i}\theta^{\prime\prime}}\alpha_3\rangle $}
        & $({D_2},{D_6},{D_9}), ({D_1},{D_5},{D_{10}}), ({D_3},{D_7},{D_{12}}), ({D_4},{D_8},{D_{11}})$ & $Z_\text{Charlie}^\text{P}$ & \\
        
\multirow{16}{*}{$|e^{7\textrm{i}\theta}\alpha _1\rangle |e^{4\textrm{i}\theta ^{\prime}}\alpha_2\rangle |e^{\pm2\textrm{i}\theta^{\prime\prime}}\alpha_3\rangle$}
        & $({D_2},{D_5},{D_{10}}), ({D_1},{D_6},{D_9}),    ({D_3},{D_8},{D_{11}}), ({D_4},{D_7},{D_{12}})$ &  $Z_\text{Bob}^\text{P}$                                 & \\
        & $({D_2},{D_5},{D_9}),    ({D_1},{D_6},{D_{10}}), ({D_3},{D_8},{D_{12}}), ({D_4},{D_7},{D_{11}})$ &  $Z_\text{Alice}^\text{P}$                               & \\
        & $({D_2},{D_7},{D_{10}}), ({D_1},{D_8},{D_9}),    ({D_3},{D_6},{D_{11}}), ({D_4},{D_5},{D_{12}})$ &  $Z_\text{Bob}^\text{S}$                                 & \\
        & $({D_2},{D_7},{D_9}),    ({D_1},{D_8},{D_{10}}), ({D_3},{D_6},{D_{12}}), ({D_4},{D_5},{D_{11}})$ &  $Z_\text{Bob}^\text{S}$,   $Z_\text{Charlie}^\text{P}$  & \\
        & $({D_2},{D_8},{D_{10}}), ({D_1},{D_7},{D_9}),    ({D_3},{D_5},{D_{11}}), ({D_4},{D_6},{D_{12}})$ &  $Z_\text{Bob}^\text{P}$,   $Z_\text{Bob}^\text{S}$      & \\
        & $({D_2},{D_8},{D_9}),    ({D_1},{D_7},{D_{10}}), ({D_3},{D_5},{D_{12}}), ({D_4},{D_6},{D_{11}})$ &  $Z_\text{Alice}^\text{P}$, $Z_\text{Bob}^\text{S}$      & \\
        & $({D_3},{D_6},{D_{10}}), ({D_4},{D_5},{D_9}),    ({D_2},{D_7},{D_{11}}), ({D_1},{D_8},{D_{12}})$ &  $Z_\text{Alice}^\text{S}$                               & \\
        & $({D_3},{D_6},{D_9}),    ({D_4},{D_5},{D_{10}}), ({D_2},{D_7},{D_{12}}), ({D_1},{D_8},{D_{11}})$ &  $Z_\text{Alice}^\text{S}$,  $Z_\text{Charlie}^\text{P}$ & \\
        & $({D_3},{D_5},{D_{10}}), ({D_4},{D_6},{D_9}),    ({D_1},{D_7},{D_{12}}), ({D_2},{D_8},{D_{11}})$ &  $Z_\text{Alice}^\text{S}$,  $Z_\text{Bob}^\text{P}$     & \\
        & $({D_3},{D_5},{D_9}),    ({D_4},{D_6},{D_{10}}), ({D_1},{D_7},{D_{11}}), ({D_2},{D_8},{D_{12}})$ &  $Z_\text{Alice}^\text{P}$,  $Z_\text{Alice}^\text{S}$   & \\
        & $({D_3},{D_7},{D_{10}}), ({D_4},{D_8},{D_9}),    ({D_2},{D_6},{D_{11}}), ({D_1},{D_5},{D_{12}})$ &  $Z_\text{Charlie}^\text{S}$ & \\
        & $({D_3},{D_7},{D_9}),    ({D_4},{D_8},{D_{10}}), ({D_2},{D_6},{D_{12}}), ({D_1},{D_5},{D_{11}})$ &  $Z_\text{Charlie}^\text{P}$, $Z_\text{Charlie}^\text{S}$ & \\
        & $({D_3},{D_8},{D_{10}}), ({D_4},{D_7},{D_9}),   ({D_2},{D_5},{D_{11}}),  ({D_1},{D_6},{D_{12}})$ &  $Z_\text{Bob}^\text{P}$,    $Z_\text{Charlie}^\text{S}$  & \\
        & $({D_3},{D_8},{D_9}),    ({D_4},{D_7},{D_{10}}), ({D_2},{D_5},{D_{12}}), ({D_1},{D_6},{D_{11}})$ &  $Z_\text{Alice}^\text{P}$,  $Z_\text{Charlie}^\text{S}$  & \\

		\hline 
        \hline
	\end{tabular}
\end{table*}

\begin{table*}[htpb]
	\centering
	\caption{Relations between the QNDs, the triggers of the single-photon detectors, and the classical feed-forward
operations used to complete the three-fusion scheme, converting both the large-scale hyper-$W$ state and the partially recyclable cases.}	\label{table3}
	\begin{tabular}{ccccc}	
		\hline 	
		\hline

$|X\rangle \langle X|$  &  Single-photon detector pairs & Feed-forward operations & Result  \\	

		\hline

\multirow{6}{*}{$|e^{9\textrm{i}\theta}\alpha_1\rangle  |e^{5\textrm{i}\theta^{\prime}}\alpha_2\rangle |\alpha_3\rangle$} 
              & $({D_2},{D_6},{D_{10}}), ({D_2},{D_6},{D_9}), ({D_2},{D_5},{D_{10}}), ({D_2},{D_5},{D_9})$ & \multirow{8}{*}{None} & \multirow{6}{*}{$|\Psi_{\text{PS}^\text{s}}\rangle$} \\

\multirow{6}{*}{$|e^{12\textrm{i}\theta}\alpha _1\rangle |e^{4\textrm{i}\theta^{\prime}}\alpha _2\rangle |\alpha _3\rangle$} 
              & $({D_2},{D_6},{D_{11}}), ({D_2},{D_6},{D_{12}}), ({D_2},{D_5},{D_{11}}), ({D_2},{D_5},{D_{12}})$ & & \multirow{6}{*}{$|\Psi_{\text{PS}^\text{s},\text{PR}^\text{s}}\rangle$}\\

\multirow{6}{*}{$|e^{3\textrm{i}\theta}\alpha_1\rangle  |e^{2\textrm{i}\theta^{\prime}}\alpha_2\rangle |\alpha _3\rangle$} 
              & $({D_1},{D_6},{D_{10}}), ({D_1},{D_6},{D_9}), ({D_1},{D_5},{D_{10}}), ({D_1},{D_5},{D_9})$       & & \multirow{6}{*}{$|\Psi_{\text{PR}^\text{ppp},\text{PS}^\text{s}}\rangle$} \\

\multirow{6}{*}{$|e^{6\textrm{i}\theta}\alpha_1\rangle  |e^{\textrm{i}\theta^{\prime}}\alpha_2\rangle  |\alpha_3\rangle$} 
              & $({D_1},{D_6},{D_{11}}), ({D_1},{D_6},{D_{12}}), ({D_1},{D_5},{D_{11}}), ({D_1},{D_5},{D_{12}})$ & 
              & \multirow{6}{*}{$|\Psi_{\text{PR}^\text{ppp},\text{PS}^\text{s},\text{PR}^\text{s}}\rangle$} \\
              
              & $({D_3},{D_7},{D_{10}}),({D_3},{D_7},{D_9}),({D_3},{D_8},{D_{10}}),({D_3},{D_8},{D_9})$ & & \\
              & $({D_3},{D_7},{D_{11}}),({D_3},{D_7},{D_{12}}),({D_3},{D_8},{D_{11}}),({D_3},{D_8},{D_{12}})$ & & \\
              & $({D_4},{D_7},{D_{10}}),({D_4},{D_7},{D_9}),({D_4},{D_8},{D_{10}}),({D_4},{D_8},{D_9})$ & & \\
              & $({D_4},{D_7},{D_{11}}),({D_4},{D_7},{D_{12}}),({D_4},{D_8},{D_{11}}),({D_4},{D_8},{D_{12}})$ & & \\

        & & &\\

\multirow{6}{*}{$|e^{9\textrm{i}\theta}\alpha_1\rangle |e^{5\textrm{i}\theta^{\prime}}\alpha_2\rangle|\alpha_3\rangle$} 
              & $({D_2},{D_7},{D_{10}}), ({D_2},{D_7},{D_9}), ({D_2},{D_8},{D_{10}}), ({D_2},{D_8},{D_9})$ & \multirow{8}{*}{$Z_{\text{Alice}}^{\text{S}}$} & \multirow{6}{*}{$|\Psi_{\text{PS}^\text{s}} \rangle$} \\

\multirow{6}{*}{$|e^{12\textrm{i}\theta}\alpha_1\rangle |e^{4\textrm{i}\theta^{\prime}}\alpha_2\rangle|\alpha_3\rangle$} 
              & $({D_2},{D_7},{D_{11}}), ({D_2},{D_7},{D_{12}}), ({D_2},{D_8},{D_{11}}), ({D_2},{D_8},{D_{12}})$ & & \multirow{6}{*}{$|\Psi_{\text{PS}^\text{s},\text{PR}^\text{s}} \rangle$}\\

\multirow{6}{*}{$|e^{3\textrm{i}\theta}\alpha_1\rangle |e^{2\textrm{i}\theta^{\prime}}\alpha_2\rangle|\alpha _3\rangle$} 
              & $({D_1},{D_7},{D_{10}}), ({D_1},{D_7},{D_9}), ({D_1},{D_8},{D_{10}}), ({D_1},{D_8},{D_9})$       & & \multirow{6}{*}{$|\Psi_{\text{PR}^\text{ppp},\text{PS}^\text{s}}\rangle$} \\

\multirow{6}{*}{$|e^{6\textrm{i}\theta}\alpha_1\rangle |e^{\textrm{i}\theta^{\prime}}\alpha_2\rangle|\alpha_3\rangle$} 
              & $({D_1},{D_7},{D_{11}}), ({D_1},{D_7},{D_{12}}), ({D_1},{D_8},{D_{11}}), ({D_1},{D_8},{D_{12}})$ & 
              & \multirow{6}{*}{$|\Psi_{\text{PR}^\text{ppp},\text{PS}^\text{s},\text{PR}^\text{s}}\rangle$} \\
              
              & $({D_3},{D_6},{D_{10}}), ({D_3},{D_6},{D_9}),   ({D_3},{D_5},{D_{10}}), ({D_3},{D_5},{D_9})$    & & \\
              & $({D_3},{D_6},{D_{11}}), ({D_3},{D_6},{D_{12}}),({D_3},{D_5},{D_{11}}), ({D_3},{D_5},{D_{12}})$ & & \\
              & $({D_4},{D_6},{D_{10}}), ({D_4},{D_6},{D_9}),   ({D_4},{D_5},{D_{10}}), ({D_4},{D_5},{D_9})$    & & \\
              & $({D_4},{D_6},{D_{11}}), ({D_4},{D_6},{D_{12}}),({D_4},{D_5},{D_{11}}), ({D_4},{D_5},{D_{12}})$ & & \\

        & & &\\

\multirow{6}{*}{$|e^{10\textrm{i}\theta}\alpha_1\rangle |e^{3\textrm{i}\theta^{\prime}}\alpha_2\rangle |e^{\pm2\textrm{i}\theta^{\prime\prime}}\alpha_3\rangle$} 
              & $({D_2},{D_6},{D_{10}}),({D_2},{D_6},{D_9}),({D_2},{D_6},{D_{11}}),({D_2},{D_6},{D_{12}})$ & \multirow{8}{*}{None} & \multirow{6}{*}{$|\Psi_{\text{PS}^\text{p},\text{PR}^\text{ss}} \rangle$} \\

\multirow{6}{*}{$|e^{12\textrm{i}\theta}\alpha_1\rangle |e^{2\textrm{i}\theta^{\prime}}\alpha_2\rangle |e^{\pm\textrm{i}\theta^{\prime\prime}}\alpha_3\rangle$} 
              & $({D_2},{D_7},{D_{10}}),({D_2},{D_7},{D_9}),({D_2},{D_7},{D_{11}}),({D_2},{D_7},{D_{12}})$ & & \multirow{6}{*}{$|\Psi_{\text{PS}^\text{p},\text{PR}^\text{sss}}\rangle$}\\

\multirow{6}{*}{$|e^{5\textrm{i}\theta}\alpha_1\rangle  |e^{3\textrm{i}\theta^{\prime}}\alpha_2\rangle |e^{\pm2\textrm{i}\theta^{\prime\prime}}\alpha_3\rangle$} 
              & $({D_3},{D_6},{D_{10}}),({D_3},{D_6},{D_9}),({D_3},{D_6},{D_{11}}),({D_3},{D_6},{D_{12}})$ & & \multirow{6}{*}{$|\Psi_{\text{PS}^\text{p},\text{PR}^\text{pss}}\rangle$} \\

\multirow{6}{*}{$|e^{8\textrm{i}\theta}\alpha_1\rangle  |e^{2\textrm{i}\theta^{\prime}}\alpha_2\rangle |e^{\pm\textrm{i}\theta^{\prime\prime}}\alpha_3\rangle$} 
              & $({D_3},{D_7},{D_{10}}),({D_3},{D_7},{D_9}),({D_3},{D_7},{D_{11}}),({D_3},{D_7},{D_{12}})$ & & \multirow{6}{*}{$|\Psi_{\text{PS}^\text{p},\text{PR}^\text{pss}}\rangle$} \\
              
\multirow{6}{*}{$|e^{8\textrm{i}\theta}\alpha_1\rangle  |e^{2\textrm{i}\theta^{\prime}}\alpha_2\rangle |e^{\pm2\textrm{i}\theta^{\prime\prime}}\alpha_3\rangle$} 
              & $({D_1},{D_5},{D_{10}}),({D_1},{D_5},{D_9}),({D_1},{D_5},{D_{11}}),({D_1},{D_5},{D_{12}})$ & & \multirow{6}{*}{$|\Psi_{\text{PS}^\text{p},\text{PR}^\text{pss}}\rangle$} \\
              
              & $({D_1},{D_8},{D_{10}}), ({D_1},{D_8},{D_9}), ({D_1},{D_8},{D_{11}}), ({D_1},{D_8},{D_{12}})$ & & \\
              & $({D_4},{D_5},{D_{10}}), ({D_4},{D_5},{D_9}), ({D_4},{D_5},{D_{11}}), ({D_4},{D_5},{D_{12}})$ & & \\
              & $({D_4},{D_8},{D_{10}}), ({D_4},{D_8},{D_9}), ({D_4},{D_8},{D_{11}}), ({D_4},{D_8},{D_{12}})$ & & \\

        & & &\\

\multirow{6}{*}{$|e^{10\textrm{i}\theta}\alpha _1\rangle |e^{3\textrm{i}\theta^{\prime}}\alpha_2\rangle |e^{\pm2\textrm{i}\theta^{\prime\prime}}\alpha_3\rangle$} 
              & $({D_2},{D_5},{D_{10}}), ({D_2},{D_5},{D_9}), ({D_2},{D_5},{D_{11}}), ({D_2},{D_5},{D_{12}})$ & \multirow{8}{*}{$Z^{\text{P}}_{\text{Alice}}$} & \multirow{6}{*}{$|\Psi_{\text{PS}^\text{p},\text{PR}^\text{ss}} \rangle$} \\

\multirow{6}{*}{$|e^{12\textrm{i}\theta}\alpha _1\rangle |e^{2\textrm{i}\theta^{\prime}}\alpha_2\rangle |e^{\pm\textrm{i}\theta^{\prime\prime}}\alpha_3\rangle $} 
              & $({D_2},{D_8},{D_{10}}), ({D_2},{D_8},{D_9}), ({D_2},{D_8},{D_{11}}), ({D_2},{D_8},{D_{12}})$ & & \multirow{6}{*}{$|\Psi_{\text{PS}^\text{p},\text{PR}^\text{sss}}\rangle$}\\

\multirow{6}{*}{$|e^{5\textrm{i}\theta}\alpha _1\rangle  |e^{3\textrm{i}\theta^{\prime}}\alpha_2\rangle |e^{\pm2\textrm{i}\theta^{\prime\prime}}\alpha_3\rangle$} 
              & $({D_3},{D_5},{D_{10}}), ({D_3},{D_5},{D_9}), ({D_3},{D_5},{D_{11}}), ({D_3},{D_5},{D_{12}})$ & & \multirow{6}{*}{$|\Psi_{\text{PS}^\text{p},\text{PR}^\text{pss}}\rangle$} \\
              
\multirow{6}{*}{$|e^{8\textrm{i}\theta}\alpha _1\rangle  |e^{2\textrm{i}\theta^{\prime}}\alpha_2\rangle |e^{\pm\textrm{i}\theta^{\prime\prime}}\alpha_3\rangle$} 
              & $({D_3},{D_8},{D_{10}}), ({D_3},{D_8},{D_9}), ({D_3},{D_8},{D_{11}}), ({D_3},{D_8},{D_{12}})$ & & \multirow{6}{*}{$|\Psi_{\text{PS}^\text{p},\text{PR}^\text{pss}}\rangle$} \\

\multirow{6}{*}{$|e^{8\textrm{i}\theta}\alpha _1\rangle |e^{2\textrm{i}\theta^{\prime}}\alpha_2\rangle  |e^{\pm2\textrm{i}\theta^{\prime\prime}}\alpha_3\rangle$} 
              & $({D_1},{D_6},{D_{10}}), ({D_1},{D_6},{D_9}), ({D_1},{D_6},{D_{11}}), ({D_1},{D_6},{D_{12}})$ & & \multirow{6}{*}{$|\Psi_{\text{PS}^\text{p},\text{PR}^\text{pss}}\rangle$} \\
              
              & $({D_1},{D_7},{D_{10}}), ({D_1},{D_7},{D_9}), ({D_1},{D_7},{D_{11}}),  ({D_1},{D_7},{D_{12}})$ & & \\
              & $({D_4},{D_6},{D_{10}}), ({D_4},{D_6},{D_9}), ({D_4},{D_6},{D_{ 11}}), ({D_4},{D_6},{D_{12}})$ & & \\
              & $({D_4},{D_7},{D_{10}}), ({D_4},{D_7},{D_9}), ({D_4},{D_7},{D_{ 11}}), ({D_4},{D_7},{D_{12}})$ & & \\
              
		\hline
		\hline

	\end{tabular}
\end{table*}

\section{Estimateingthe performances of the hyperfusion schemes} \label{sec4}

\subsection{Success probabilities of the schemes} \label{sec41}

In this section, we evaluate the success probabilities of our two-fusion and three-fusion protocols.
Based on Sec. \ref{sec2}, one can see that the success probability of the desired two-fused large-scale hyper-$W$ state $|W_{n+m-2}\rangle_P \otimes |W_{n+m-2}\rangle_S$ is
\begin{eqnarray}\label{eq35}
P_\text{S}=\frac{(n+m-2)^2}{n^2m^2}.
\end{eqnarray}
The failure probability of the two-fusion protocol (the garbage state) is
\begin{eqnarray}\label{eq36}
P_\text{F}=\frac{1}{n^2 m^2}.
\end{eqnarray}
The success probability of the two-fusion partially successful state and partially recyclable states is given by
\begin{eqnarray}\label{eq36}
P_\text{PS,PR}=1-\frac{(n+m-2)^2}{n^2m^2}-\frac{1}{n^2 m^2}.
\end{eqnarray}

Based on Sec. \ref{sec3}, we can see that the success probability of the desired large-scale hyper-$W$ state $|W_{n+m+t-3}\rangle_P \otimes |W_{n+m+t-3}\rangle_S$ is 
\begin{eqnarray}\label{eq37}
P_\text{S}^{\prime}=\frac{(n+m+t-3)^2}{n^2m^2t^2}.
\end{eqnarray}
The failure probability of the three-fusion protocol is
\begin{eqnarray}\label{eq38}
P_\text{F}^{\prime}=\frac{1}{n^2m^2t^2}.
\end{eqnarray}
The success probability of the three-fusion partially successful state and partially recyclable states is
\begin{eqnarray}\label{eq38}
P_\text{PS,PR}^{\prime}=1-\frac{(n+m+t-3)^2}{n^2m^2t^t}-\frac{1}{n^2m^2t^2}.
\end{eqnarray}

\begin{figure}[htpb]
	\centering	\includegraphics[width=1\linewidth]{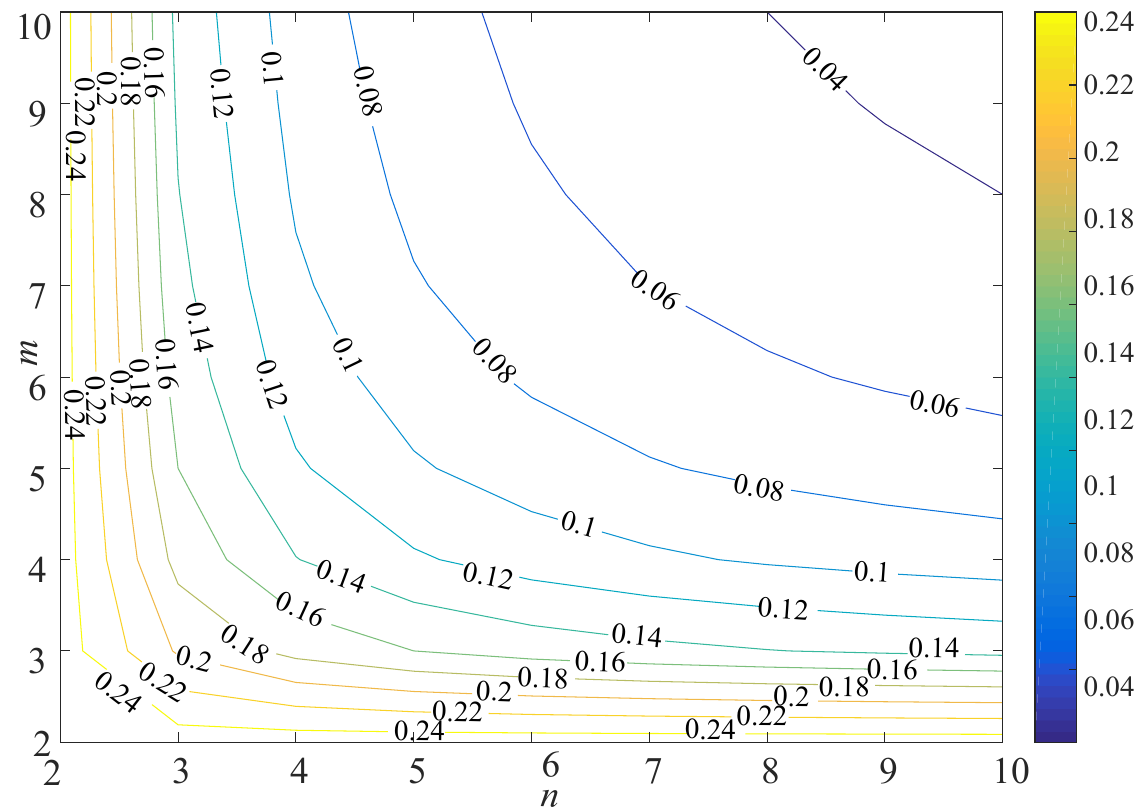}
	\caption{Success probability of the presented two-fusion protocol when $2 \le n$ and $m \le 10$.}
	\label{figure3}
\end{figure}

\begin{figure}[htpb]
	\centering
	\subfigure{
		\begin{minipage}[t]{1\linewidth}
			\includegraphics[width=1\linewidth]{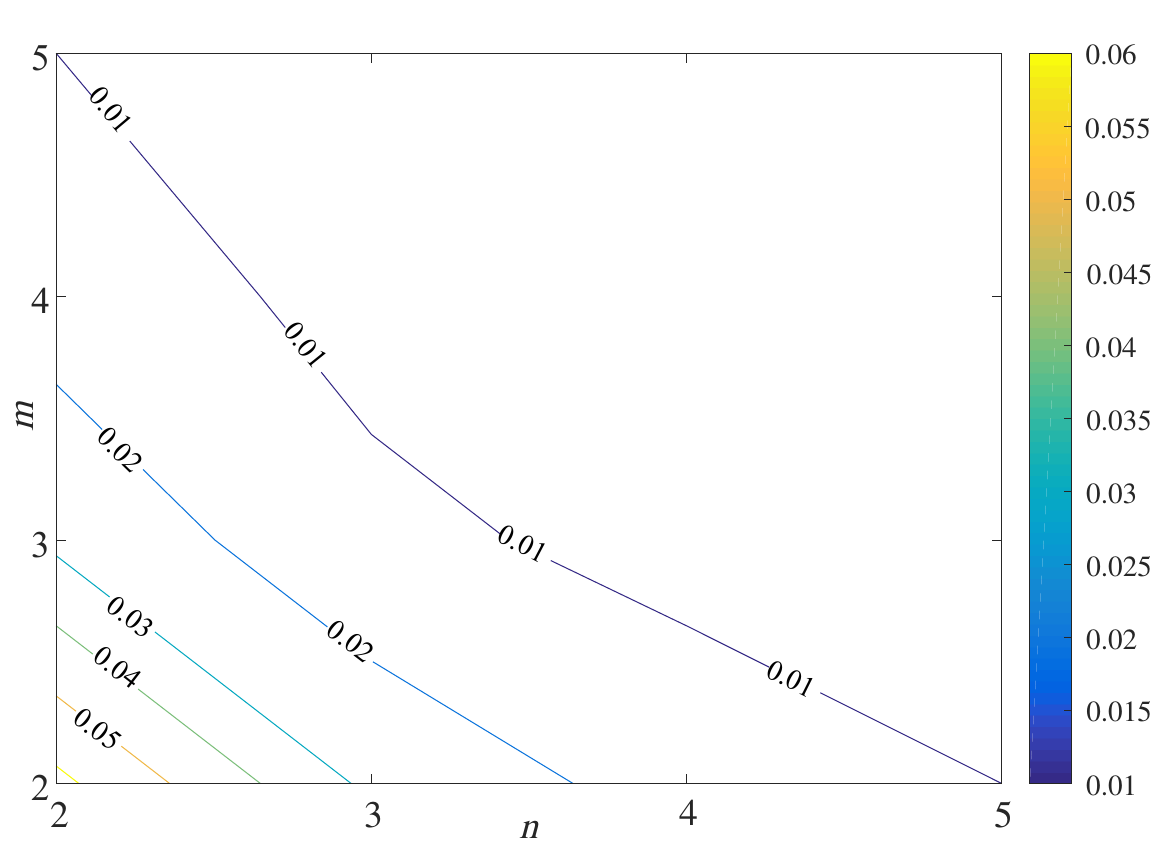}\\
			\includegraphics[width=1\linewidth]{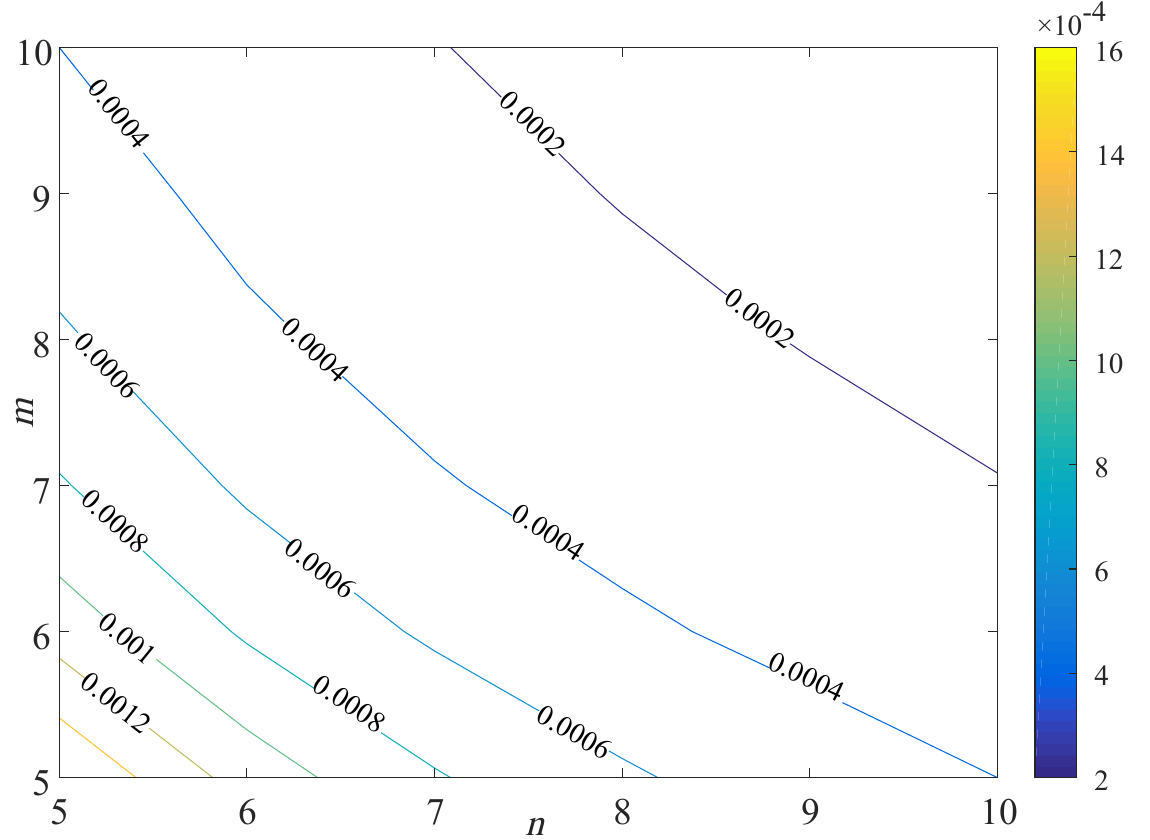}
		\end{minipage}%
	}%
	\centering
     \caption{Failure probability of the presented two-fusion protocol when $2 \le n$ and $m \le 10$.}
	\label{figure4}
\end{figure}

\begin{figure}[htpb]
	\centering
	\subfigure{
		\begin{minipage}[t]{1\linewidth}
			\includegraphics[width=1\linewidth]{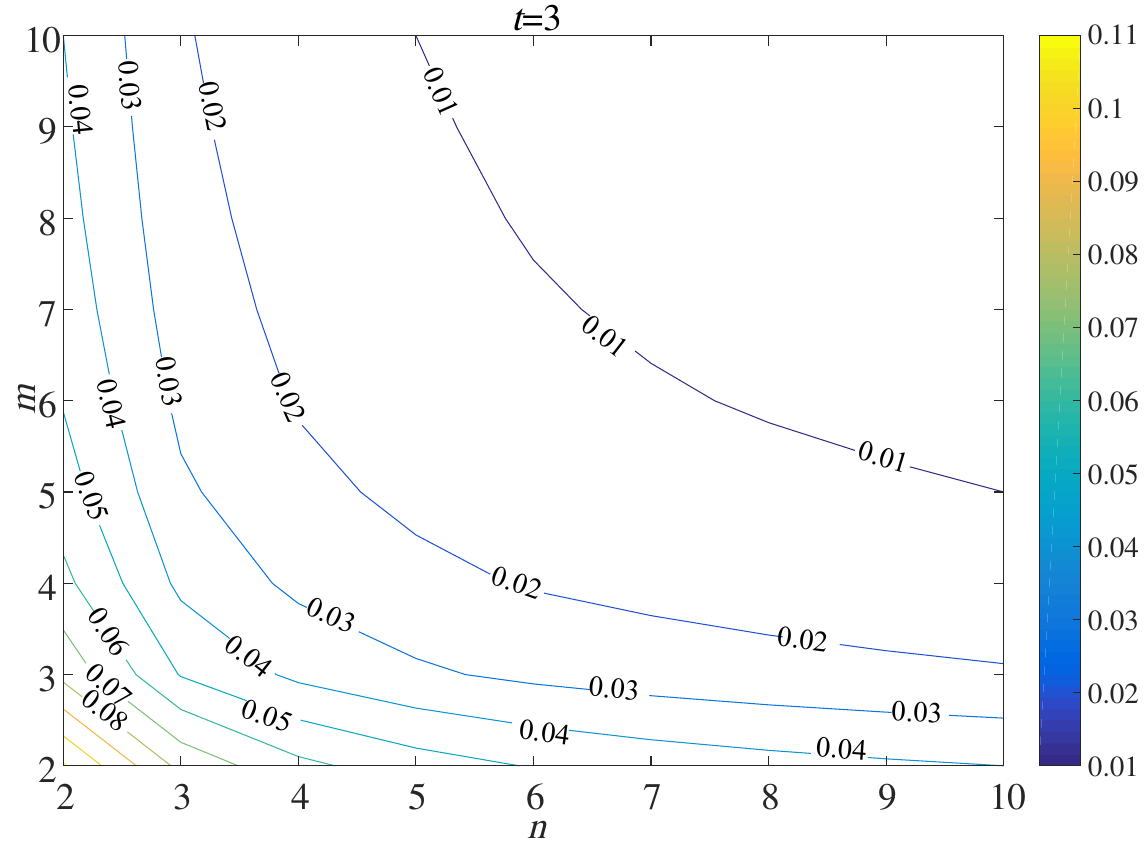}
			\includegraphics[width=1\linewidth]{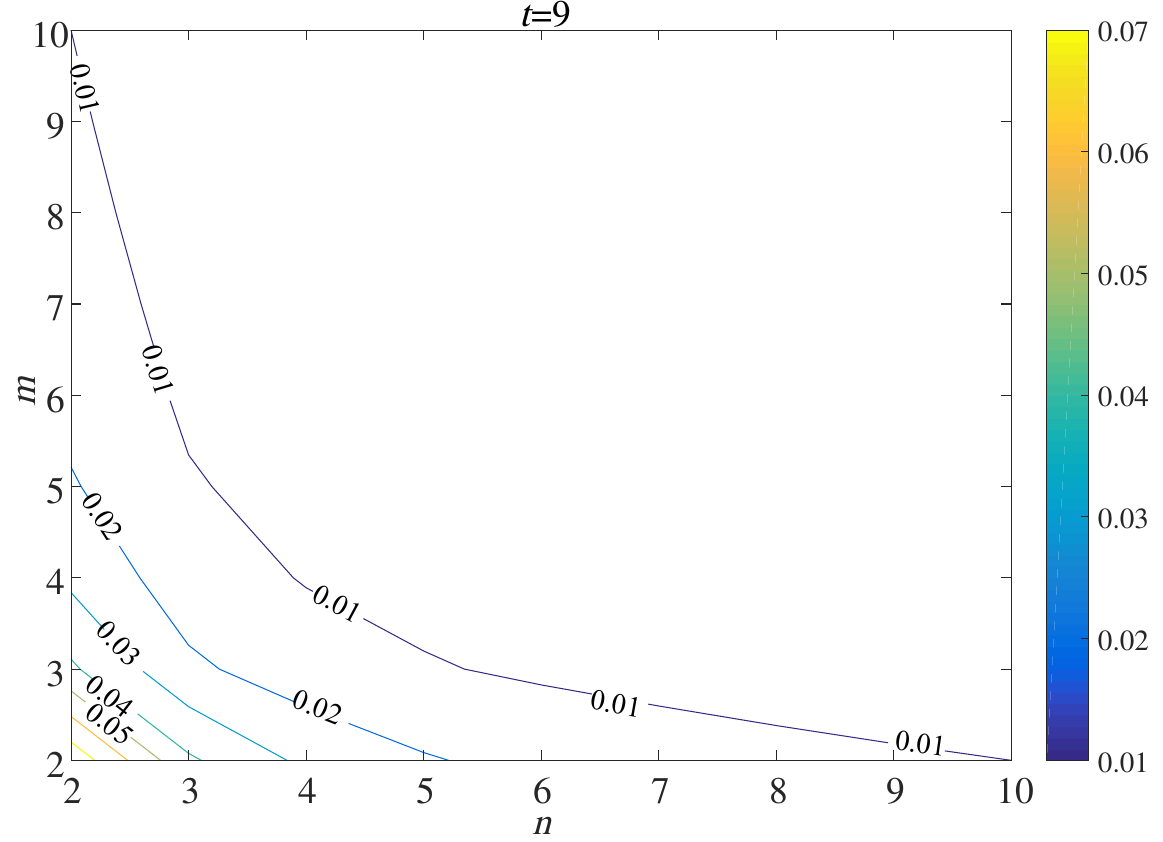}
		\end{minipage}%
	}%
	\centering
		\caption{Success probability of the presented three-fusion protocol when $2 \le n$, $m \le 10$, and $t=3, 9$.}
	\label{figure5}
\end{figure}

\begin{figure*}[htpb]
	\centering
	\subfigure{
		\begin{minipage}[t]{1\linewidth}
			\includegraphics[width=0.46\linewidth]{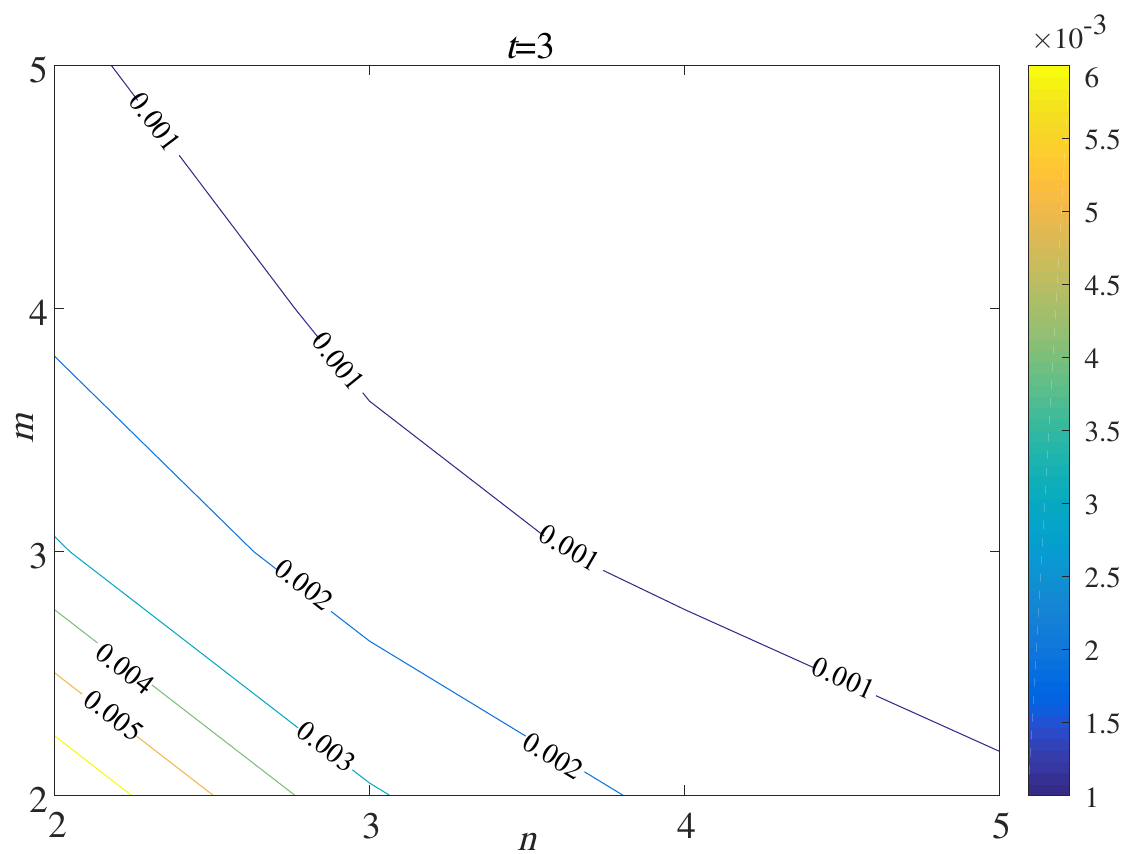}
			\includegraphics[width=0.46\linewidth]{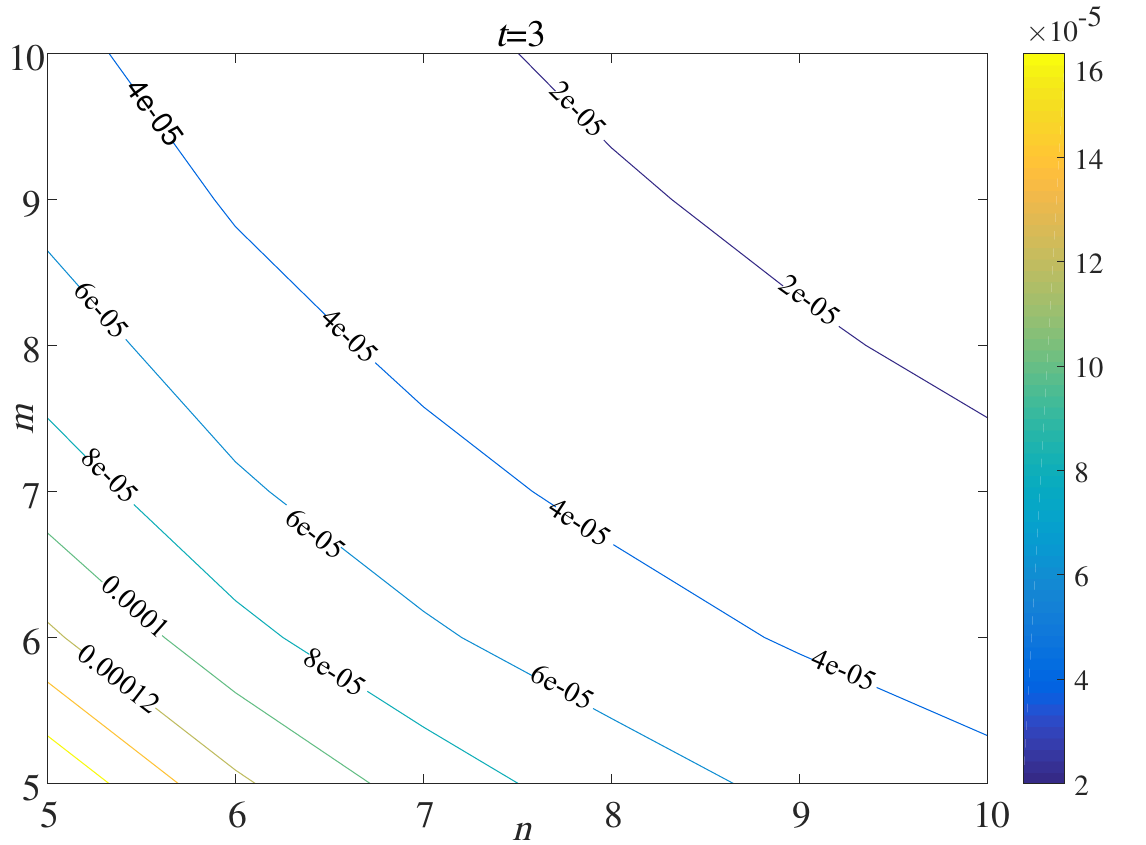}
		\end{minipage}%
	}%
	
	\subfigure{
		\begin{minipage}[t]{1\linewidth}
			\includegraphics[width=0.46\linewidth]{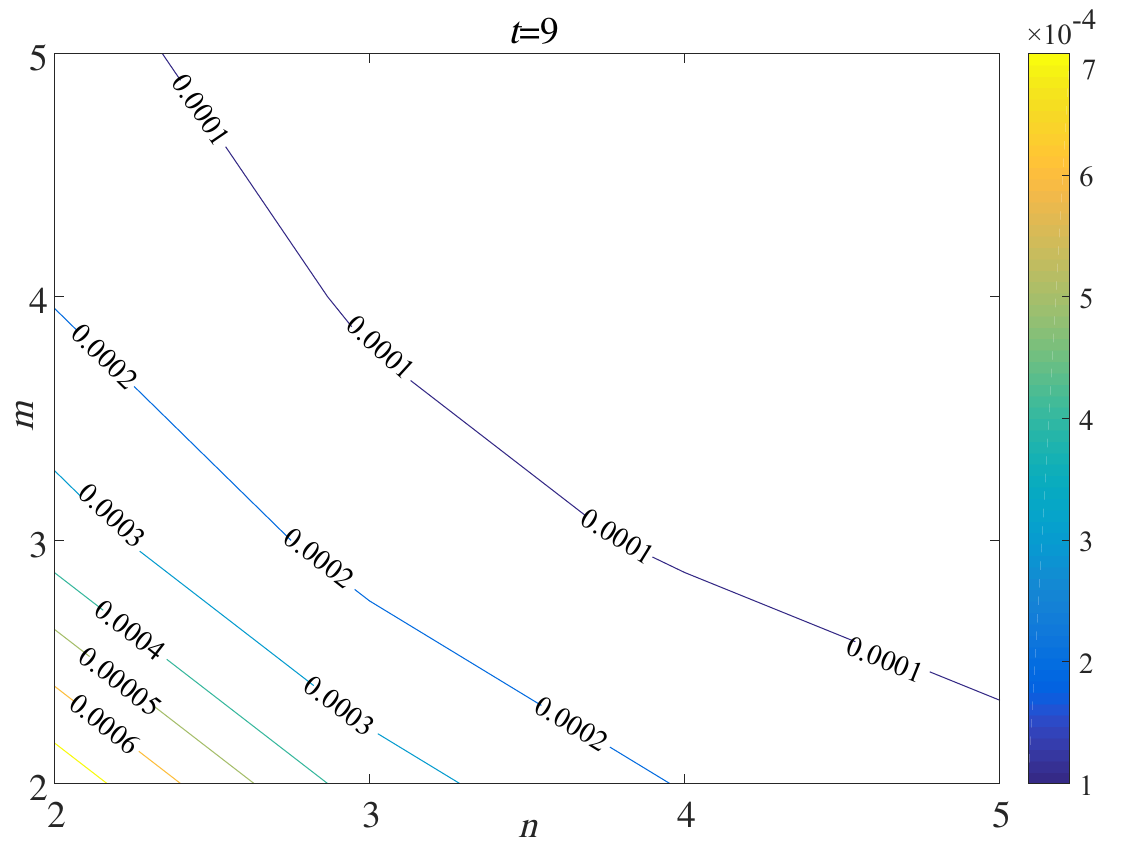}
			\includegraphics[width=0.46\linewidth]{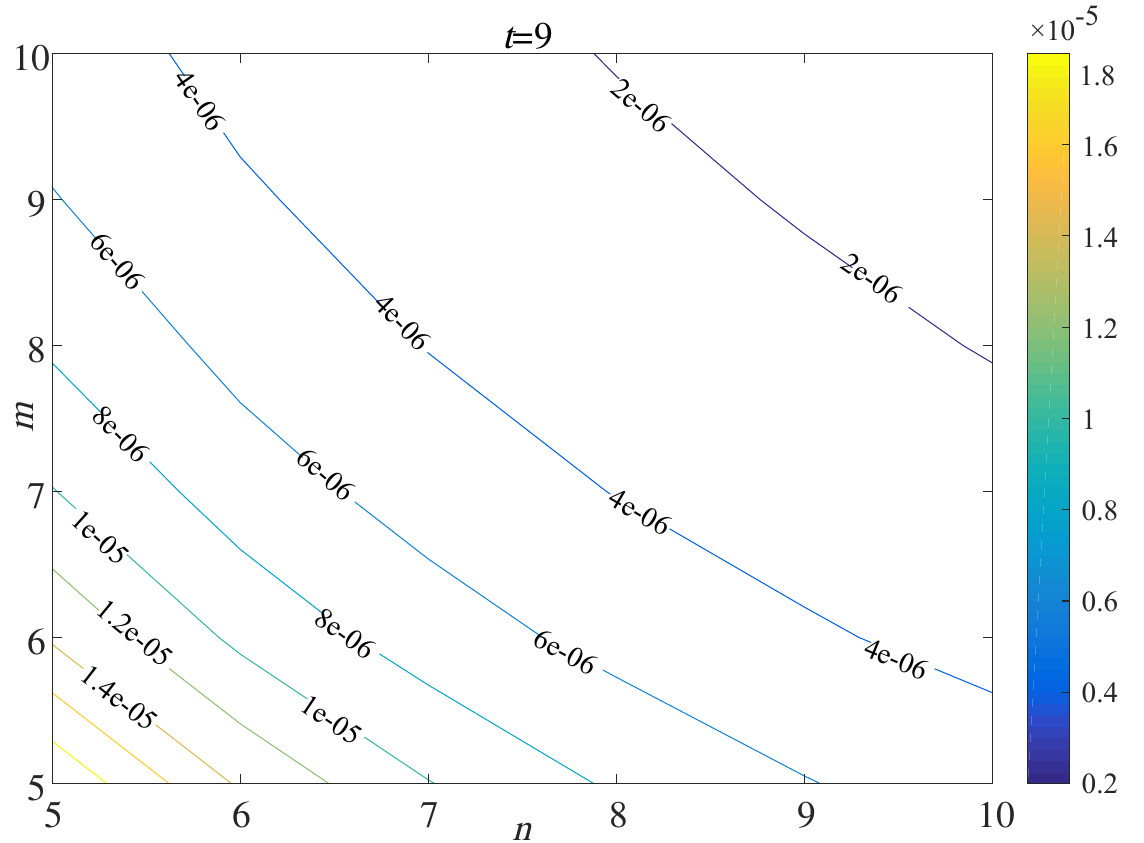}
		\end{minipage}
	}%
	\centering
	\caption{Failure probability of the presented three-fusion protocol when $2 \le n$, $m \le 10$, and $t=3,9$.}
	\label{figure6}
\end{figure*}

Based on Figs. \ref{figure3}-\ref{figure6}, we can see how the performances of our two-fusion and three-fusion schemes are affected by adjusting $n$ and $m$.
It is clear that the failure probability is much less than the success probability.
Hence, our schemes are relatively efficient.

\subsection{Imperfections arising from linear optics} \label{sec42}

BSs, PBSs, HWPs, and single-photon detectors are the necessary ingredients in our two hyper-$W$ fusion schemes.
Hence, in  practical applications, the imperfections arising from these linear optical elements will cause the real output states to deviate from their ideal counterpart, and they will also  slightly reduce the fidelity of the fused hyper-$W$ states.
The transformations of the imperfect PBS can be given by \cite{imperfection}
\begin{eqnarray}\label{eq39}
\begin{aligned}
U_{\text{PBS}}|H\rangle = & \frac{1}{\sqrt{1+r}}[(\cos\theta-\sqrt{r}\sin\theta) |H\rangle \\&
                                                -(\sin\theta+\sqrt{r}^{*}\cos\theta) |V\rangle],
\end{aligned}
\end{eqnarray}
\begin{eqnarray}\label{eq40}
\begin{aligned}
U_{\text{PBS}}|V\rangle = & \frac{1}{\sqrt{1+r}}[(\sin\theta+\sqrt{r}\cos\theta) |H\rangle \\&
                                                +(\cos\theta-\sqrt{r}^{*}\sin\theta) |V\rangle],
\end{aligned}
\end{eqnarray}
where $\theta$ is the deviation of the mirror mounts and $r$ is the polarization extinction ratio of the PBS.

The imperfection in the transmission ratio of the balanced BS, denoted by $\epsilon$ modifies the transformations of the 50:50 BS as
\begin{eqnarray}\label{eq41}
\begin{aligned}
U_{\text{BS}}|\textbf{0}\rangle_S = \frac{1}{\sqrt{\epsilon^2+2\epsilon+2}}[(1+\epsilon)|\textbf{0}\rangle_S + |\textbf{1}\rangle_S],
\end{aligned}
\end{eqnarray}
\begin{eqnarray}\label{eq42}
\begin{aligned}
U_{\text{BS}}|\textbf{1}\rangle_S = \frac{1}{\sqrt{\epsilon^2+2\epsilon+2}}[|\textbf{0}\rangle_S - (1+\epsilon) |\textbf{1}\rangle_S].
\end{aligned}
\end{eqnarray}

In additional to the BS and PBS, the detector dark count (1.2\%), two-photon events (1.2\%), and imperfections in the polarization HWP plate (0.8\%) also effect the fidelity. We believe that these issues will be mitigated by future developments.

\subsection{Imperfections arising from cross-Kerr nonlinearities} \label{sec43}

\begin{figure*}[htbp]
	\centering
	\includegraphics[width=0.68\linewidth]{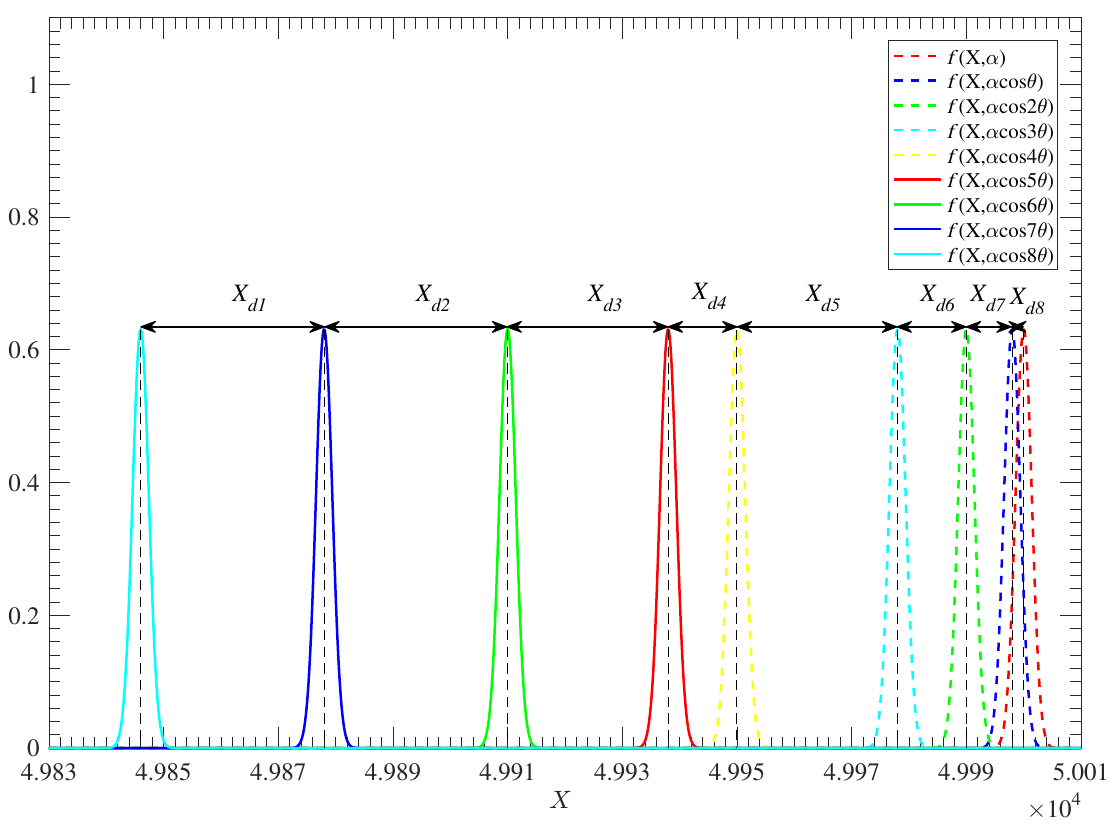}
	\caption{Gaussian function curves of
        $f(X,\alpha )$ (red dashed line),
		$f(X,\alpha \cos\theta)$ (blue dashed line),
        $f(X,\alpha \cos2\theta)$ (green dashed line),
        $f(X,\alpha \cos3\theta)$ (cyan dashed line),
        $f(X,\alpha \cos4\theta)$ (yellow dashed  line),
        $f(X,\alpha \cos5\theta)$ (red solid line),
        $f(X,\alpha \cos6\theta)$ (green solid line),
        $f(X,\alpha \cos7\theta)$ (blue solid line), and
        $f(X,\alpha \cos8\theta)$ (cyan solid line). Here,
        $\alpha = 2500$ and $\theta = 0.01$ are used.}
	\label{figure7}
\end{figure*}

\begin{figure*}[htbp]
	\centering
	\includegraphics[width=0.68\linewidth]{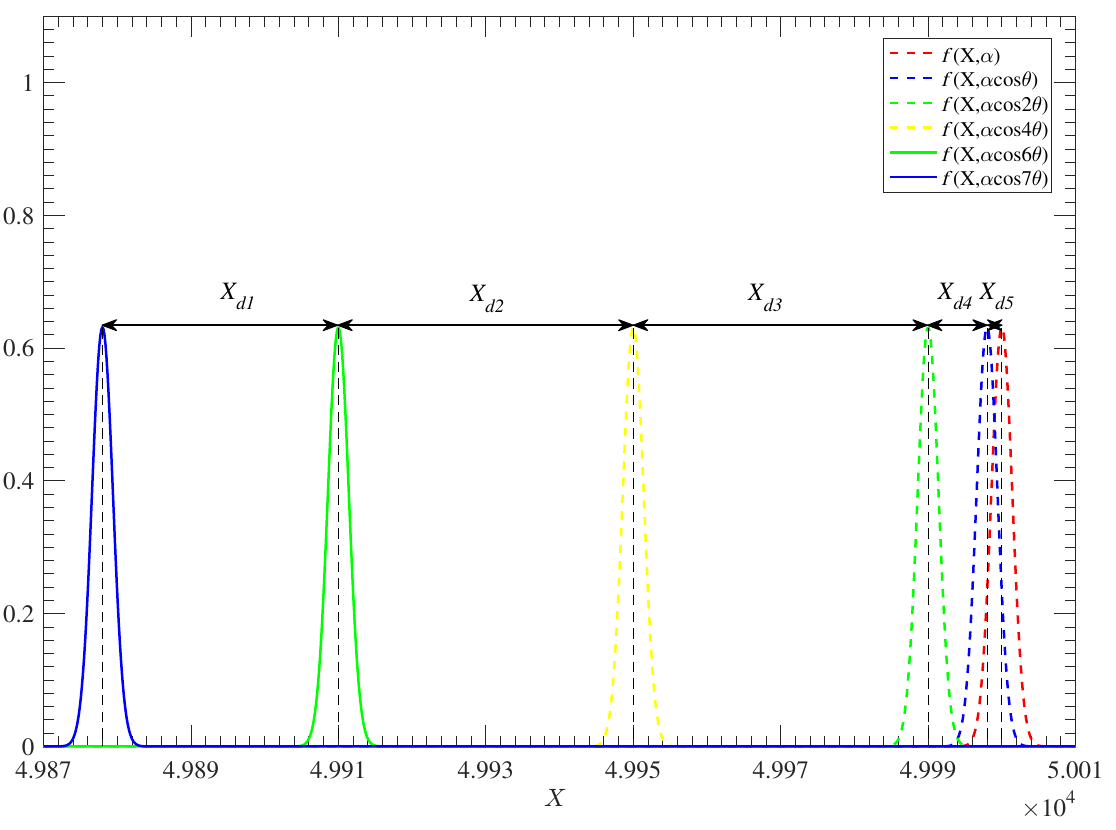}
	\caption{Gaussian function curves of
        $f(X,\alpha )$ (red dashed line),
		$f(X,\alpha \cos\theta )$ (blue dashed line),
        $f(X,\alpha \cos2\theta )$ (green dashed line),
        $f(X,\alpha \cos4\theta )$ (yellow  dashed line),
        $f(X,\alpha \cos6\theta )$ (green solid line), and
        $f(X,\alpha \cos7\theta )$ (blue solid line).
        Here $\alpha = 2500$ and $\theta = 0.01$ are used.}
	\label{figure8}
\end{figure*}

Cross-Kerr nonlinearities are the key elements of our hyper-$W$ fusion schemes. Hence, the error arising from the $X$-homodyne measurements should be taken into account.
To clarify our description, we will take the two-fused scheme shown in Sec. \ref{sec2} as an example.
In Fig. \ref{figure1},  nine distinguishable measurement outcomes (i.e., the peaks of Gaussian curves occurring at 0, $\theta$, $2\theta$, $3\theta$, $4\theta$, $5\theta$, $6\theta$, $7\theta$, and $8\theta$) of the coherent state $|\alpha\rangle$ are required. As shown in Figs. \ref{figure7}-\ref{figure8}, the phase-shit error arises mainly from the partial overlap between
$f(X,\alpha)$ and $f(X,\alpha\cos\theta)$.
This occurs because the distances between
$f(X,\alpha\cos2\theta)$ and $f(X,\alpha\cos3\theta)$,
$f(X,\alpha\cos3\theta)$ and $f(X,\alpha\cos4\theta)$,
$f(X,\alpha\cos4\theta)$ and $ f(X,\alpha\cos5\theta)$,
$f(X,\alpha\cos5\theta)$ and $ f(X,\alpha\cos6\theta)$,
$f(X,\alpha\cos6\theta)$ and $ f(X,\alpha\cos7\theta)$, and
$f(X,\alpha\cos7\theta)$ and $ f(X,\alpha\cos8\theta)$ are much larger than the distance between
$f(X,\alpha)$ and $f(X,\alpha\cos\theta)$.
Here, $f(X,\alpha \cos n\theta)= (2\pi)^{-\frac{1}{4}}\exp{[-\frac{1}{4}(x-2\cos n\theta)^2]}$, and $X_{dn}$ denotes the distance between the peaks of two adjacent Gaussian curves.
Using electromagnetically induced transparency \cite{EIT}, the magnitude of the natural cross-Kerr nonlinearity $\theta$ can be enhanced to around $10^{-2}$.

When the error of the homodyne measurement is taken into account, the success probability of the homodyne measurement $P_{\text{suc}}$ in Fig. \ref{figure1} is given by
\begin{eqnarray}\label{p_suc}
	{P_{\text{suc}}}(\alpha,\theta,\gamma t) = {(1-\frac{1}{2}\text{erfc}[\frac{{{e^{\frac{-\gamma t}{2}\alpha (1 - \cos\theta)}}}}{\sqrt2}])},
\end{eqnarray}
while success probability ${P_{\text{suc}}}_{'}$ of the homodyne measurement in Fig. \ref{figure2} is given by
\begin{eqnarray}\label{p_suc}
	{P_{\text{suc}}}_{'}(\alpha ,\theta ,\gamma t) = {(1 - \frac{1}{2}\text{erfc}[\frac{{{e^{\frac{-\gamma t}{2}\alpha (1-\cos\theta)}}}}{\sqrt2}])^3},
\end{eqnarray}
where $\gamma$ is the decay constant.

\begin{figure*}[htpb]
	\centering
	\subfigure{
		\begin{minipage}[t]{1\linewidth}
			\includegraphics[width=0.46\linewidth]{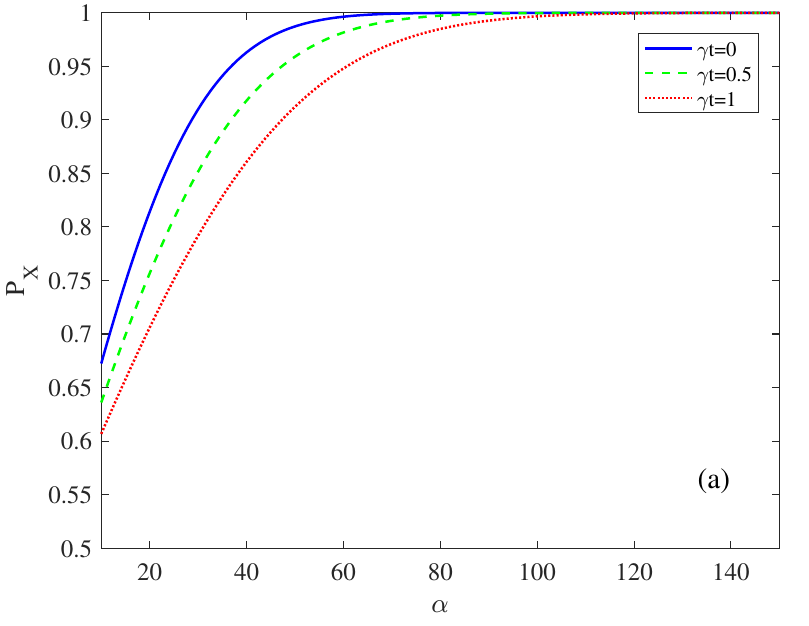}
			\includegraphics[width=0.46\linewidth]{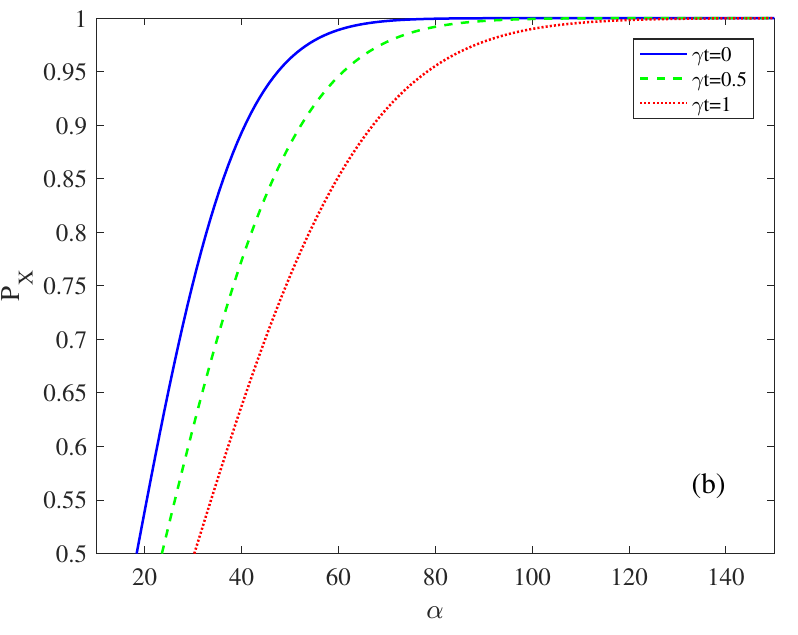}
		\end{minipage}%
	}
	\centering
   \caption{Success probabilities of the homodyne measurements as a function of the amplitude $\alpha$ under the conditions 
   $\gamma t=0$ (blue solid line),
   $\gamma t=0.5$ (green dashed line) and
   $\gamma t=1$ (red dotted line). (a) $P_X$ for Fig. \ref{figure1}. (b) $P_X$ for Fig. \ref{figure2}. Here, $\theta =0.3$ is used.}
	\label{figure9}
\end{figure*}

Based on Fig. \ref{figure9}, we see that a high success probability $P_X$ of the $X$-homodyne measurement  can be achieved with a larger amplitude $\alpha$ of the coherent state and a smaller dissipation coefficient $\gamma t$.

\section{conclusion } \label{sec5}

In conclusion, we have proposed two practical hyperfusion schemes for hyper-$W$ states via weak cross-Kerr nonlinearities.
The computation qubits are encoded in the polarization and spatial DOFs of single-photon systems.
The two-fusion protocol can fuse one $n$-photon hyper-$W$ state and one $m$-photon hyper-$W$ state into an $(n+m-2)$-photon hyper-$W$ state, 
while the three-fusion protocol can fuse one $n$-photon hyper-$W$ state, one $m$-photon hyper-$W$ state, and one $t$-photon hyper-$W$ state into an $(n+m+t-3)$-photon hyper-$W$ state.
Unlike previous fusion schemes, which are limited to single-DOF systems such as polarization or spin,
our fused $W$ states are hyperentangled in both polarization and spatial DOFs.
Our schemes are more efficient, producing only one output garbage state.
The constructions of the two-fused and three-fused hyper-$W$ states employ PBSs, BSs, HWPs, single-photon detectors, and cross-kerr nonlinearities;
quantum conditional controlled gates and additional single photons are not required.
Designing schemes to fuse four, five, or more hyper-$W$ states simultaneously, as well as developing schemes to fuse hyper-$W$ states without qubits loss, will be addressed in our future work.

\medskip

\section*{Acknowledgment}
This work is funded by the National Natural Science Foundation of China under Grant No. 62371038, the Science Research Project of Hebei Education Department under Grant No. QN2025054, and the Fundamental Research Funds for the Central Universities under Grant No. FRF-TP-19-011A3.

\medskip

\section*{Appendix}

In the second step of Sec. \ref{sec2}, an X-quadrature homodyne measurement is performed on the coherent state.
In detail, if the quantum nondemolition detection (QND) result is $|\alpha\rangle$, the system will be projected into the state
\begin{eqnarray}\label{eq3}
	\begin{split}
		|\Phi_{e^{\text{i}0\theta}}\rangle =\;& \frac{1}{nm}|(n-1)_H\rangle_P |(m-1)_H\rangle_P |(n-1)_{a_0}\rangle_S \\&\otimes |(m-1)_{b_0}\rangle_S |1_{V_{a_1}}\rangle_S |1_{V_{b_1}}\rangle.
	\end{split}
\end{eqnarray}

If the detection result is $|e^{\text{i}\theta}\alpha\rangle$, the system will be projected into the state
\begin{eqnarray}\label{eq4}
	\begin{split}
		|\Phi_{e^{\text{i}\theta}}\rangle =\;&
		\frac{\sqrt{m-1}}{nm}|(n-1)_H\rangle_P |W_{m-1}\rangle_P \\&\otimes   |(n-1)_{a_0}\rangle_S |(m-1)_{b_0}\rangle_S |1_{V_{a_1}}\rangle |1_{H_{a_1}}\rangle\\&
		+ \frac{\sqrt{n-1}}{nm}|W_{n-1}\rangle_P |(m-1)_H\rangle_P \\&\otimes |(n-1)_{a_0}\rangle_S |(m-1)_{b_0}\rangle_S |1_{H_{b_1}}\rangle |1_{V_{b_1}}\rangle.
	\end{split}
\end{eqnarray}

If the detection result is $|e^{\text{i}2\theta}\alpha\rangle$, the system will be projected into the state
\begin{eqnarray}\label{eq5}
	\begin{split}
		|\Phi_{e^{\text{i}2\theta}}\rangle =\;&
		\frac{\sqrt{n-1}\sqrt{m-1}}{nm}|W_{n-1}\rangle_P |W_{m-1}\rangle_P \\&\otimes |(n-1)_{a_0} \rangle_S |(m-1)_{b_0} \rangle_S |1_{H_{b_1}}\rangle |1_{H_{a_1}}\rangle.
	\end{split}
\end{eqnarray}

If the detection result is $|e^{\text{i}3\theta}\alpha\rangle$, the system will be projected into the state
\begin{eqnarray}\label{eq6}
	\begin{split}
		|\Phi_{e^{\text{i}3\theta}}\rangle =\;&
		\frac{\sqrt{m-1}}{nm}|(n-1)_H\rangle_P |(m-1)_H\rangle_P \\&\otimes  |(n-1)_{a_0}\rangle_S |W_{m-1}\rangle_S      |1_{V_{a_1}}\rangle |1_{V_{b_0}}\rangle\\&
		+\frac{\sqrt{n-1}}{nm}|(n-1)_H\rangle_P |(m-1)_H\rangle_P \\&\otimes |W_{n-1}\rangle_S     |(m-1)_{b_0}\rangle_S  |1_{V_{a_0}}\rangle |1_{V_{b_1}}\rangle.
	\end{split}
\end{eqnarray}

If the detection result is $|e^{\text{i}4\theta}\alpha\rangle$, the system will be projected into the state
\begin{eqnarray}\label{eq7}
	\begin{split}
		|\Phi_{e^{\text{i}4\theta}}\rangle =\; &
		\frac{m-1}{nm}                 |(n-1)_H\rangle_P  |W_{m-1}\rangle_P  \\&\otimes |(n-1)_{a_0}\rangle_S   |W_{m-1}\rangle_S      |1_{V_{a_1}}\rangle_S  |1_{H_{a_0}}\rangle \\ &
		+\frac{\sqrt{n-1}\sqrt{m-1}}{nm}|(n-1)_H\rangle_P  |W_{m-1}\rangle_P \\&\otimes |W_{n-1}\rangle_S      |(m-1)_{b_0}\rangle_S  |1_{V_{a_0}}\rangle_S  |1_{H_{a_1}}\rangle\\&
		+\frac{\sqrt{n-1}\sqrt{m-1}}{nm}|W_{n-1}\rangle_P  |(m-1)_H\rangle_P \\&\otimes |(n-1)_{a_0}\rangle_S  |W_{m-1}\rangle_S      |1_{H_{b_1}}\rangle_S |1_{V_{b_0}}\rangle\\&
		+\frac{n-1}{nm}                 |W_{n-1}\rangle_P  |(m-1)_H\rangle_P \\&\otimes |W_{n-1}\rangle_S      |(m-1)_{b_0}\rangle_S  |1_{H_{b_0}}\rangle_S |1_{V_{b_1}}\rangle.
	\end{split}
\end{eqnarray}

If the detection result is $|e^{\text{i}5\theta}\alpha\rangle$, the system will be projected into the state
\begin{eqnarray}\label{eq8}
	\begin{split}
		|\Phi_{e^{\text{i}5\theta}}\rangle =\; &
		 \frac{(n-1)\sqrt{m-1}}{nm} |W_{n-1}\rangle_P |W_{m-1}\rangle_P \\&\otimes  |W_{n-1}\rangle_S    |(m-1)_{b_0}\rangle_S |1_{H_{b_0}}\rangle |1_{H_{a_1}}\rangle\\&
		+\frac{\sqrt{n-1}(m-1)}{nm} |W_{n-1}\rangle_P |W_{m-1}\rangle_P \\&\otimes |(n-1)_{a_0}\rangle_S |W_{m-1}\rangle_S     |1_{H_{b_1}}\rangle |1_{H_{a_0}}\rangle.
	\end{split}
\end{eqnarray}

If the detection result is $|e^{\text{i}6\theta}\alpha\rangle$, the system will be projected into the state
\begin{eqnarray}\label{eq9}
	\begin{split}
		|\Phi_{e^{\text{i}6\theta}}\rangle =\;& \frac{\sqrt{n-1}\sqrt{m-1}}{nm} |(n-1)_H\rangle_P |(m-1)_H\rangle_P \\&\otimes |W_{n-1}\rangle_S |W_{m-1}\rangle_S |1_{V_{a_0}}\rangle |1_{V_{b_0}}\rangle.
	\end{split}
\end{eqnarray}

If the detection result is $|e^{\text{i}7\theta}\alpha\rangle$, the system will be projected into the state
\begin{eqnarray}\label{eq10}
	\begin{split}
		|\Phi_{e^{\text{i}7\theta}}\rangle =\; &
		 \frac{\sqrt{n-1}(m-1)}{nm}|(n-1)_H\rangle_P |W_{m-1}\rangle_P \\&\otimes |W_{n-1}\rangle_S |W_{m-1}\rangle_S  |1_{V_{a_0}}\rangle |1_{H_{a_0}}\rangle\\&
		+\frac{(n-1)\sqrt{m-1}}{nm}|W_{n-1}\rangle_P |(m-1)_H\rangle_P \\&\otimes |W_{n-1}\rangle_S |W_{m-1}\rangle_S |1_{H_{b_0}}\rangle |1_{V_{b_0}}\rangle.
	\end{split}
\end{eqnarray}

If the detection result is $|e^{\text{i}8\theta}\alpha\rangle$, the system will be projected into the state
\begin{eqnarray}\label{eq11}
	\begin{split}
		|\Phi_{e^{\text{i}8\theta}}\rangle =\;& \frac{(n-1)(m-1)}{nm} |W_{n-1}\rangle_P |W_{m-1}\rangle_P \\&\otimes |W_{n-1}\rangle_S |W_{m-1}\rangle_S |1_{H_{b_0}}\rangle |1_{H_{a_0}}\rangle.
	\end{split}
\end{eqnarray}

\end{document}